\documentclass[pra,amssymb,amsmath,superscriptaddress,twocolumn,footinbib,floatfix]{revtex4-2}
\usepackage{braket}
\usepackage{float}
\usepackage{tikz}
\usepackage{multirow}
\usepackage{booktabs}
\usepackage[export]{adjustbox}
\usepackage{graphicx}
\usepackage{afterpage}
\usepackage[page,toc]{appendix}
\usepackage{amsthm}
\usepackage{dsfont}
\usepackage{hyperref}
\usepackage{chngcntr}
\hypersetup{
	colorlinks=true,
	linkcolor=blue,
	filecolor=magenta,      
	urlcolor=magenta,
	citecolor=teal
 }
\usepackage{xcolor}
\usepackage[normalem]{ulem}
\usepackage{soul}
\DeclareMathOperator{\Tr}{Tr}

\DeclareMathOperator{\ketup}{\ket{\uparrow}}
\DeclareMathOperator{\ketdown}{\ket{\downarrow}}

\newcommand{\sys}{\mathrm{sys}}
\newcommand{\env}{\mathrm{env}}
\newcommand{\itn}{\mathrm{int}}
\newcommand{\outprod}[1]{\ket{#1}\!\!\bra{#1}}
\newcommand{\Sket}[1]{\ket{#1}\!\rangle}

\newcommand{\proj}[2]{\ket{#1}\!\!\bra{#2}}
\usepackage{dcolumn}
\usepackage{bm}
\usepackage{blkarray}
\usepackage{xparse}
\usepackage{mathtools}
\usepackage{subfiles}
\usepackage{xcolor}

\DeclareMathOperator{\rhoTLS}{{\rho}_{\mathrm{sys}}^{(\mathrm{TL})}}

\begin{document}
	
	\preprint{APS/123-QED}
	
 \title{Phonon-Induced Decoherence in Color-Center Qubits}
	\author{Prajit Dhara}%
 \email[]{prajitd@arizona.edu}
	\affiliation{Wyant College of Optical Sciences, The University of Arizona, Tucson, AZ 85721}
	\affiliation{NSF-ERC Center for Quantum Networks, The University of Arizona, Tucson, AZ 85721}
	\author{Saikat Guha}
 \email[]{saikat@arizona.edu}
	\affiliation{Wyant College of Optical Sciences, The University of Arizona, Tucson, AZ 85721}
	\affiliation{NSF-ERC Center for Quantum Networks, The University of Arizona, Tucson, AZ 85721}
	
	\begin{abstract}
Electron spin states of solid-state defects such as Nitrogen- and Silicon-vacancy {\em color centers} in diamond are a leading quantum-memory candidate for quantum communications and computing. Via open-quantum-systems modeling of spin-phonon coupling---the major contributor of decoherence---at a given temperature, we derive the time dynamics of the density operator of an electron-spin qubit. We use our model to corroborate experimentally-measured decoherence rates. We further derive the temporal decay of distillable entanglement in spin-spin entangled states heralded via photonic Bell-state measurements. Extensions of our model to include other decoherence mechanisms, e.g., undesired hyperfine couplings to the neighboring nuclear-spin environment, will pave the way to a rigorous predictive model for engineering artificial-atom qubits with desirable properties.
	\end{abstract}
	
	\maketitle
	\section{Introduction}
	\label{sec:intro}
	
	Many quantum information processing tasks, esp., in quantum computing and in quantum repeaters for long-distance quantum communication, rely on systems that can serve as quantum memories, i.e., can store states of qubits for extended periods of time, and processors, i.e., allow for quantum logic gates to be performed on the stored qubits. 
	Of the many candidate technologies for quantum information storage, solid state quantum memories based on color center defects in diamond have emerged as leading candidates~\cite{Nguyen2019-gv,Schmidgall2018-cv,Rugar2020-ol,Pingault2017-yo,Bradley2019-em}, especially because these systems promise long coherence times, high-fidelity single-qubit gates, efficient qubit-photon interfaces, and in-principle highly-scalable realizations. Integration of these defect centers into nanophotonic waveguides promises a scalable pathway for the development of fault-tolerant quantum repeaters---either with the spin qubits themselves acting as quantum memories~\cite{Dhara2021}, or where the spin qubits are used to produce photonic cluster states that act like quantum memories~\cite{Pant2017}---which would serve as the backbone of the quantum internet. 
	
	Candidate spin vacancies in diamond comprise of two major classes -- the nitrogen vacancy (NV) or a group IV vacancy (G4V), where the candidate defect atom is either silicon (Si), germanium (Ge), tin (Sn) or lead (Pb)~\cite{Thiering2018-vp,Gali2019-sl,Sipahigil2016-de,Debroux2021-mw,Trusheim2020-ko,Wang2021-qp,Iwasaki2015-kh,Ruf2021-ye,Trusheim2019-ck,Parker2023-yh}. While NVs have been used in pioneering demonstrations of verifiable shared entanglement generation~\cite{Pompili2021-bp,Hermans2022-zl}, problems with low photon coupling efficiency and inherent susceptibility to electronic charge noise leaves room for improvement. Of the G4Vs, the negatively charged silicon vacancy (SiV) has been a prominent candidate of choice due to a multitude of properties. The inversion symmetric vacancy has a negligible susceptibility to electronic charge noise, with near-deterministic implantation of defects promising scalable device manufacturing~\cite{Wan2020-op}. Demonstrations of memory enhanced quantum communications~\cite{Bhaskar2020-kv} and long coherence times approaching 2 sec~\cite{Stas2022-yw} are a few key reasons for increased interest in these systems. Phonon coupling is the major source of decoherence in these systems at `high' temperatures~\cite{Jahnke2015-no}, necessitating the operation of typical experiments at 150 mK or lower, something which must be mitigated for scalable network deployments.  Heavier defects such as the tin-vacancy (SnV) serve as promising candidates for higher temperature operation, however, a deep understanding of phonon decoherence is crucial for the effective utilization  of this class of vacancies. Previous studies into the effect of spin-phonon coupling have quantified and corroborated theoretically-predicted decoherence rates with experimental predictions. However, a complete dynamical characterization of the underlying {\em quantum state} through a master equation or similar state evolution maps has been lacking in the literature. A recent article~\cite{Harris2023-aq} in this area, relied on first principles calculation of the phonon coupling mechanism and rate.
	
	In this article, we address the spin-phonon coupling effect for G4Vs, and derive a Born-Markov master equation for the complete quantum state. We use the derived equation to characterize the time evolution of the quantum states, both of a single spin and two entangled spins. Furthermore, we address the effect of decoherence on shared entanglement generation over a network using the `midpoint swap' architecture~\cite{Barrett2005-sm,Dhara2022,Hermans2022-je}. We show that with the consideration of network latency, strict conditions and/or limitations are imposed on the quality of the final entangled qubit pair, which we quantify using a lower bound on its distillable entanglement per copy.
	
	The article is organized as follows. We review the electronic structure, underlying system Hamiltonians and state energy level structures in Sec.~\ref{sec:electronic_structure}. Interaction of the vacancy energy manifold with a phonon bath is analyzed to compose the master equation in Sec.~\ref{sec:master_eq}. We quantify the effect of decoherence on single and entangled spin qubit states in Sec.~\ref{sec:evol_analysis}, and obtain the temperature dependent decoherence rates. Section~\ref{sec:entanglement_decohere} analyzes the quality of entanglement generated over a single quantum link with the effect of decoherence and communication latency; we also address the potential questions regarding multi-party entanglement. We conclude our study with outlooks to future work in Sec.~\ref{sec:conclusion}.
	
	\section{Electronic Configuration of Group-IV Vacancies in Diamond}
	\label{sec:electronic_structure}
	The wavelength of the principal optical transition in G4Vs---closely related to its ground and excited state splitting---is defect-atomic-species dependent. Besides interfaces for single photon generation, G4Vs are promising platforms for quantum information processing, which is aided by its split ground level electronic manifold. 
	
\begin{figure}[htbp]
		\centering
		\includegraphics[width=\linewidth]{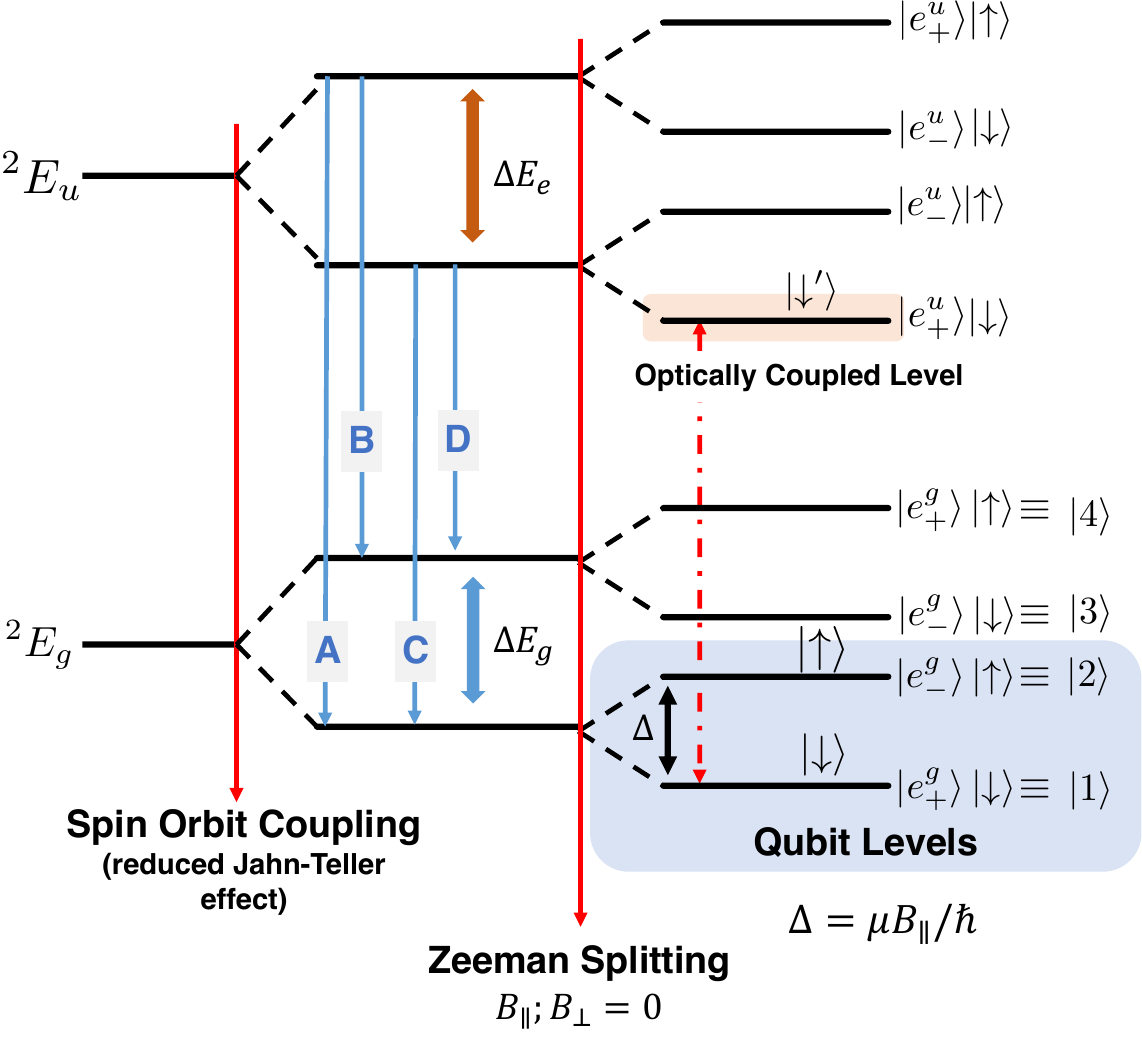}
		\caption{Electronic level structure of a Group IV vacancy center in diamond. Optical transitions from a typical spectra (shown in inset) are marked using blue lines. Level splitting occurs under various interactions (marked by red lines). The levels highlighted in blue form the qubit system used for tasks in quantum information processing.  }
		\label{fig:elevel}
	\end{figure}

	The ground level manifold (represented in the literature~\cite{Hepp2014-fa,Hepp2014-ws} and in Fig.~\ref{fig:elevel} as $ ^{2}E_g $) is a four-fold degenerate energy level with an orbital and a spin degree of freedom. The Hilbert space for $^{2}E_g$ can be expressed as $ \mathcal{H}=\mathcal{H}_{\mathrm{orbital}}\otimes \mathcal{H}_{\mathrm{spin}} $, where $\mathcal{H}_{\mathrm{orbital}}$ is the orbital subsystem and $ \mathcal{H}_{\mathrm{spin}}$ is the spin subsystem, both of which are two dimensional Hilbert spaces. The basis states of $ \mathcal{H}_{\mathrm{orbital}}$ are expressed as $ \ket{e^g_+} $ and $ \ket{e^g_-} $, whereas for $ \mathcal{H}_{\mathrm{spin}} $, they are $ \ketup $ and $ \ketdown $. Since we limit the discussion to the ground state manifold, we may henceforth drop the $g$ superscript for the orbital basis states, i.e.\ $\ket{e^g_\pm}\equiv\ket{e_\pm}$. However, fine structure splitting in the spectra of these systems are not attributable to these bare levels i.e., they must account for additional interactions. Theoretical and experimental~\cite{Thiering2018-vp,Hepp2014-fa} studies have attributed the splitting in these systems to three major interactions, that we shall review and formulate in formal notation subsequently.
	
	\emph{Spin-orbit coupling} --- Relativistic interaction of the electronic orbital with the nuclear potential of the defect atom is the cause of spin orbit coupling. Normally a rotation invariant interaction, the crystal field of the host diamond breaks the symmetry for G4Vs to yield the interaction Hamiltonian~\cite{Hepp2014-fa,Yu2010-fc}, 
	\begin{subequations}
		\begin{align}
			{H}_{\mathrm{SO}}&=-\frac{\lambda_{\mathrm{SO}}}{2} (-\proj{e_+}{e_+}+ \proj{e_-}{e_-})\otimes (\outprod{\uparrow}-\outprod{\downarrow}) \\
			&=+\frac{\lambda_{\mathrm{SO}}}{2} \cdot  \hat{Z}_{\mathrm{orb.}}\otimes \hat{Z}_{\mathrm{spin}},
		\end{align}
	\end{subequations}
	where $ \lambda_{\mathrm{SO}} $ is the spin-orbit coupling strength with $ \hat{Z}_{\mathrm{orb.}} = (\proj{e_+}{e_+}-\proj{e_-}{e_-}) $ and $ \hat{Z}_{\mathrm{spin}}= \outprod{\uparrow}-\outprod{\downarrow} $.
	
	\emph{Jahn-Teller interaction} --- This effect introduces distortion of the electronic orbitals due to an asymmetric potential, leading to orbital energy shifts~\cite{Abtew2011-lz,OBrien1993-xm}. The Jahn-Teller effect, is less prominent than spin-orbit coupling, and is a spin-independent interaction with an interaction Hamiltonian of the form,
	\begin{subequations}
		\begin{align}
			H_{\mathrm{JT}}&=\left[\Upsilon_x \hat{Y}_{\mathrm{orb.}} -\Upsilon_y \hat{X}_{\mathrm{orb.}}\right] \otimes \hat{\mathbb{I}}_{\mathrm{spin}},
		\end{align}
	\end{subequations}
	where $ \Upsilon_x,\Upsilon_y $ represent the effective energies associated with the distorted potential along the $ x,y $ directions (in the cardinal frame) respectively, and $\hat{Y}_\mathrm{orb.}=i\proj{e_+}{e_-}-i\proj{e_-}{e_+},\hat{X}_\mathrm{orb.}=\proj{e_+}{e_-}+\proj{e_-}{e_+} $.
	
	\emph{Zeeman splitting} --- Zeeman splitting is observed when external magnetic  fields lift the spin-degeneracy of electronic defect~\cite{Condon1951-kc}. There are two distinct effects dependent on the direction of the field, which may be parallel $ (B_\parallel) $ or perpendicular $ (B_\perp) $ to the high-symmetry axis of the defect center. The parallel field yields an effective Hamiltonian, 
	\begin{align}
		H_{Z,\parallel}=\frac{\gamma_e}{2}\cdot \mathbb{I}_{\mathrm{orb.}} \otimes B_\parallel \hat{Z}_{\mathrm{spin}}.
	\end{align}
	The perpendicular field yields the effective Hamiltonian,
	\begin{align}
		H_{Z,\perp}=\frac{\gamma_e}{2}\cdot \mathbb{I}_{\mathrm{orb.}} \otimes \left[B_x \hat{X}_{\mathrm{spin}} + B_y\hat{Y}_{\mathrm{spin}}\right]. 
	\end{align}
	Here, $ \gamma_e=2\mu_B /\hbar $ and $ B_x , B_y$ are the orthogonal components of perpendicular field ($ B_\perp $),  $\hat{X}_{\mathrm{spin}}=\proj{\uparrow}{\downarrow} + \proj{\downarrow}{\uparrow}$ and $\hat{Y}_{\mathrm{spin}}=i\proj{\uparrow}{\downarrow}-i\proj{\downarrow}{\uparrow}$.
	
	For the purposes of our study, we shall focus on the final structure as a result of the joint action of these Hamiltonians. In the future sections of the paper, we consider the joint effect of these interactions, and work with a fixed basis of state vectors for the ground state manifold. Specifically, we consider the scenario where the Jahn-Teller effect is negligible ($ \Upsilon_x,\Upsilon_y \ll \lambda_\mathrm{SO} $) and the system experiences an on-axis magnetic field ($ B=B_\parallel; B_\perp=0 $)~\cite{Hepp2014-ws}. The total Hamiltonian for the defect center, our \emph{system} of interest, is therefore given as:
	\begin{align}
		H_{\mathrm{sys}}\approx H_{\mathrm{SO}}+H_{Z,\parallel}.
	\end{align}
	The energy eigenstates of $H_{\mathrm{sys}}$ are,
	\begin{align}
		\begin{split}
			\ket{e_+}\otimes\ketdown \equiv \ket{1}; \ket{e_-}\otimes\ketup \equiv \ket{2},\\
			\ket{e_-}\otimes\ketdown \equiv \ket{3}; \ket{e_+}\otimes\ketup \equiv \ket{4}.
		\end{split}
	\end{align}
	For the sake of brevity, we shall use the abbreviated representation, $ \ket{i}, i=\{1,2,3,4\} $ of the eigenstates for the subsequent sections of the paper. Readers may refer to Appendix~\ref{sec:app_elecstruc} for detailed descriptions of these level structures.

	\section{Master Equation Setup for Electron Spin-Phonon Coupling}
	\label{sec:master_eq}
	Accounting for dissipation and decoherence in qubits is crucial to evaluate the utility of these systems in practical settings. Starting with the non-dissipative description of a quantum system, ad-hoc introduction of dissipative terms will violate the canonical commutation rules~\cite{Carmichael1999-zf,Breuer2002-op} of underlying operators. Dissipation in a quantum system can be analyzed using various techniques; we consider a Born-Markov master equation based approach for the purposes of the current article.
	
	In particular, we are interested in the evolution of a system of interest, represented by the density operator $ \rho_{\mathrm{sys}} $. Given solely the Hamiltonian $ H_{\mathrm{sys}} $ with no additional interactions, the evolution of the state of the system can be solved for in either the Schr\"{o}dinger or Heisenberg formalism. However, any decoherence or dissipation requires the introduction of an environment system initialized as the state $ \rho_{\mathrm{env}} $ with the Hamiltonian $ H_{\mathrm{env}} $, and an interaction Hamiltonian $ H_{\mathrm{int}} $. The evolution of the total system, $ \rho_{\mathrm{tot}}=\rho_{\mathrm{sys}} \otimes \rho_{\mathrm{env}} $ under the total Hamiltonian $ H_{\mathrm{tot}}=H_{\mathrm{sys}}+H_{\mathrm{env}}+H_{\mathrm{int}} $ is analytically expressible; however obtaining general solutions to the state evolution is generally intractable. {Under the assumptions of initial state separability for a `large' invariant environment state (Born approximation), and memoryless evolution of the system density operator (Markov approximation)}, a Born-Markov master equation may be derived for the evolution of $\rho_{\mathrm{sys}}$.
	
	\begin{figure}[h!]
		\centering
		\includegraphics[width=0.85\linewidth]{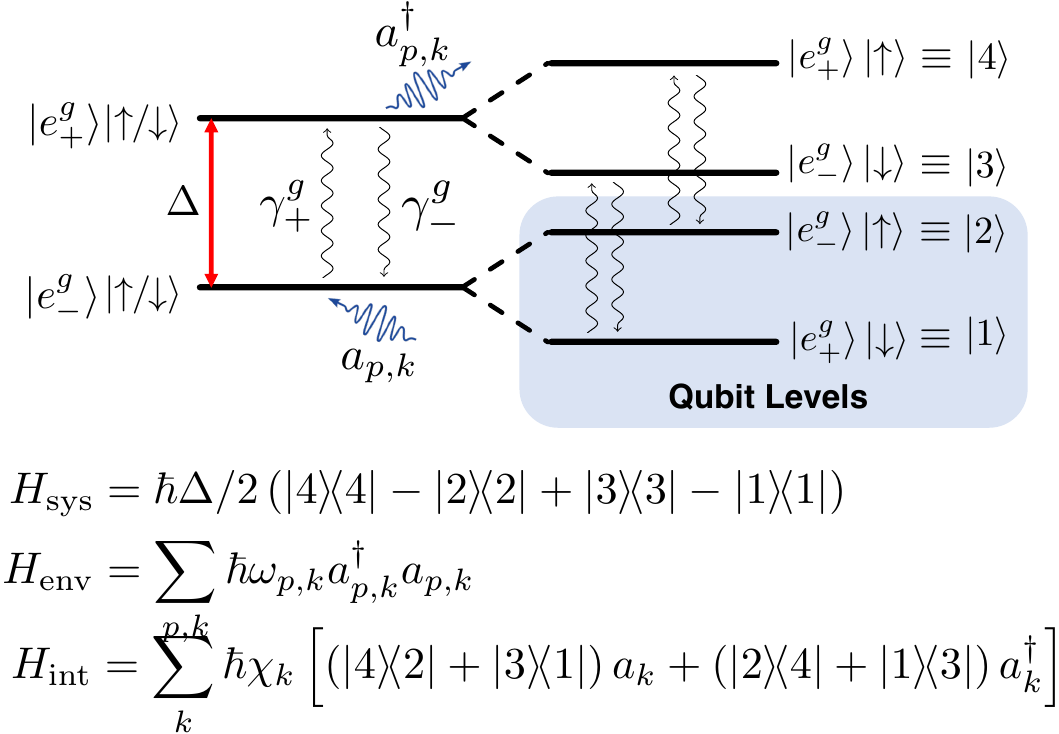}
		\caption{ Phonon coupling with split levels of the ground state manifold of the G4V center.}
		\label{fig:phonon_coupling}
	\end{figure}
	
	For the current study, we want to examine the major cause of qubit decoherence in G4Vs induced by phonon-coupling to the electronic spin-orbital levels. Experimental studies have predicted the spin-phonon coupling effect to be the major cause of decoherence at high temperatures~\cite{Jahnke2015-no}; in fact, as highlighted previously, most experiments eliminate the effect by `freezing' the phonon bath by operating at suitable temperatures (typically $ \sim $150 mK or below, for SiV color centers). 
	
	To develop an analytical model for qubit decoherence, we begin by considering the complete electronic Hamiltonian,
	\begin{align}
		H_{\mathrm{sys}}=\hbar \Delta/2\,  (\outprod{4}-\outprod{2}+\outprod{3}-\outprod{1}),
		\label{eq:sys_Ham}
	\end{align}
	where $\Delta  =\lambda_{\mathrm{SO}}$ is the frequency corresponding to the total energy splitting between $ \{\ket{1},\ket{3} \} $ and $ \{\ket{2},\ket{4}\} $ (for details see Appendix~\ref{sec:app_elecstruc} and Refs.~\cite{Hepp2014-fa,Debroux2021-mw}). The interaction of the system with phonons can be modeled by considering an phonon bath (a collection of bosonic modes) environment system with the Hamiltonian, 
	\begin{align}
		H_{\mathrm{env}}&=\sum_{p,k} \hbar \omega_{p,k} a_{p,k}^{\dagger} a_{p,k},
		\label{eq:env_Ham}
	\end{align}
	where $ a^\dagger_{p,k} (a_{p,k}) $ is the creation (annihilation) operator for phonons with polarization $ p=\{-,+\} $ and wave-vector $ k $, with $ \{\ket{l}_{p,k}; l=0,1,2,\ldots\} $ representing the Fock states of the specified mode. The initial state of the environment is a multi-mode thermal state,
	\begin{align}
		\rho_{\mathrm{env}}&=\bigotimes_{p,k} \sum_{l=0}^{\infty} \frac{\bar{n}_{k}^l}{(\bar{n}_{k}+1)^{l+1}} \outprod{l}_{k},
		\label{eq:env_state}
	\end{align}
	where $\bar{n}_k$ is the modal thermal phonon occupation number. Orbital-phonon coupling is a spin conserving transition, resulting in transition of the quantum state  between the eigen-levels $ H_{\mathrm{sys}} $ by the absorption (or emission) of a phonon. Correspondingly, the system-environment interaction Hamiltonian can be modeled as:
	\begin{align}
		H_{\mathrm{int}}=\sum_{k} \hbar \chi_{k}\left[\left(\proj{4}{2}+\proj{3}{1}\right)a_{p, k}+\left(\proj{2}{4}+\proj{1}{3}\right) a_{p,k}^{\dagger}\right],
		\label{eq:int_Ham}
	\end{align}
	where $ \chi_k $ is the interaction strength for a single-phonon absorption to, or emission from, the mode labeled by wave-vector $k$. For the diamond substrate, the interaction coefficient ($ \chi(\omega) $) and density of modes ($ g(\omega) $) are approximately given by $ \overline{|\chi(\omega)|^2}\approx\chi_0 \,\omega $ and $ g(\omega) \approx g_0\, \omega^2 $ respectively. The overbar denotes the average over all modes with frequency $ \omega_k=\omega  $. The proportionality constants $g_0,\chi_0$ are obtained from experimental studies~\cite{Jahnke2015-no,Carmichael1999-zf}. Under the Born-Markov approximation  {\footnote{This is valid since the collective phonon bath environment of a typical sample is `large' and unperturbed by the interaction with the spin levels.}}, the final master equation in the standard Lindblad form~\cite{Carmichael1999-zf,Breuer2002-op,Manzano2020-bi} given as, 
	\begin{widetext}
		\begin{align}
			\begin{split}
				\dot{\rho}=&-i \frac{1}{2} \omega_{A}^{\prime}\left[(\outprod{4}-\outprod{2}+\outprod{3}-\outprod{1}), \rho\right]\\
				&+\frac{\gamma}{2}(\bar{n}+1)\left(2\times  \ket{2} \!\!\braket{4|\rho|4} \!\! \bra{2} -\outprod{4}\rho-\rho \outprod{4}\right) 
				+\frac{\gamma}{2} \bar{n}\left(2 \times \ket{4} \!\!\braket{2|\rho|2} \!\! \bra{4}-\outprod{2} \rho-\rho \outprod{2} \right)\\
				&+\frac{\gamma}{2}(\bar{n}+1)\left(2\times  \ket{1} \!\!\braket{3|\rho|3} \!\! \bra{1} -\outprod{3}\rho-\rho \outprod{3}\right) 
				+\frac{\gamma}{2} \bar{n}\left(2 \times \ket{3} \!\!\braket{1|\rho|1} \!\! \bra{3}-\outprod{1} \rho-\rho \outprod{1} \right),
			\end{split}
			\label{eq:ME_SiV}
		\end{align}
	\end{widetext}
	where $  \bar{n}=\bar{n}(\Delta,T)= {e^{-\hbar \Delta / k_{B} T}}/({1-e^{-\hbar \Delta / k_{B} T}})$ and $ \gamma =2\pi g(\Delta)\cdot|\chi (\Delta)|^2=2\pi g_0 \chi_0 \Delta^3 $.  The energy separation is modified by a specific amount $ \omega'_A=\Delta+2\Delta'+\Delta_{\mathrm{Lamb}}  $, where the $ \Delta' $ is a temperature-dependent shift and $ \Delta_{\mathrm{Lamb}} $ is the normal Lamb shift. These modifications to the energy splitting arises from quantum vacuum fluctuations and manifests in the environment correlation integrals~\cite{Carmichael1999-zf}. Appendix~\ref{app:deriv_ME} has further details.
	
	Before we proceed with a numerical simulation of the system, we draw some insights on the derived master equation. In particular, we can focus on the dynamics of a two level system (TLS) coupled to a bosonic thermal environment~\cite{Carmichael1999-zf}. The system, environment and interaction Hamiltonians for this system are given by:
	\begin{align}
		\begin{split}
			H_{\mathrm{sys}}&=\hbar \omega_{A} (\outprod{e}-\outprod{g}), \\
			H_{\mathrm{env}}&= \sum_{m} \hbar \omega _m a_m^\dagger a_m, \,{\text{and}} \\
			H_{\mathrm{int}}& = \sum_{m} \hbar \chi_m (\proj{e}{g}a_m+\proj{g}{e} a_m^\dagger ),
		\end{split}
	\end{align}
	where $ \{\ket{e},\ket{g}\} $ are the excited and ground state levels of the TLS separated by energy of $ \hbar \omega_A  $. The bosonic environment is a collection of modes labeled by $ m $, with the corresponding  creation (annihilation) operators  given by $ a_m(a_m^\dagger) $. The interaction between the TLS and the bosonic mode is coupled by the interaction strength coefficient $ \chi_m $. A commonly analyzed situation is the evolution of the system density operator (denoted by $ \rhoTLS$) when the environment is in a collective thermal state (similar to Eq.~\eqref{eq:env_state}, with mode index $ m $ replacing  $ \{p,k\} $). Under the Born-Markov approximation, the master equation governing this evolution~\cite{Carmichael1999-zf} is given by:
	{\small
		
		\begin{align}
			\begin{split}
				\dot{\rho}_{\mathrm{sys}}^{(\mathrm{TL})}=&-i \omega_A^{\prime}/2 \left[\outprod{e}-\outprod{g}, \rhoTLS \right]\\
				&+\frac{\Gamma}{2}(\bar{n}+1)\left(2 \ket{g} \!\!\braket{e|\rhoTLS|e} \!\! \bra{g} -\{\outprod{e}, \rhoTLS\}\right) \\
				&+\frac{\Gamma}{2} \bar{n}\left(2 \times \ket{e} \!\!\braket{g|\rhoTLS|g} \!\! \bra{e}-\{\outprod{g}, \rhoTLS\} \right), 
			\end{split}
			\label{eq:ME_TLS}
		\end{align}
		
	}
	\noindent where $\{A,B\}=AB+BA$ is the anti-commutator brackets, $\bar{n}=\bar{n}(\omega_A,T)= {e^{-\hbar \omega_A / k_{B} T}}/({1-e^{-\hbar \omega_A / k_{B} T}})$ is the thermal boson occupation of the environment mode at frequency $\omega_A$ and $ \Gamma =2\pi g(\omega_{A})\cdot|\chi (\omega_A)|^2$ is the overall coupling constant. It is easy to note that Eq.~\eqref{eq:ME_SiV} and~\eqref{eq:ME_TLS} are quite similar; the general master equation for the color center is comprised of two TLSs. Namely, the pair of levels $ \{\ket{1},\ket{3}\} $ and $ \{\ket{2},\ket{4}\} $ correspond respectively to the levels $ \{\ket{g}, \ket{e}\}$ of Eq.~\eqref{eq:ME_TLS}. {This does not mean that all the analysis valid for Eq.~\eqref{eq:ME_TLS}  necessarily caries over to the analysis of decoherence using Eq.~\eqref{eq:ME_SiV}. However, one may draw parallels between the two systems for added intuition. The TLS's decay rate (transfer from $\ket{e}\rightarrow\ket{g}$ state) of $\Gamma (\bar{n}+1)/2$ is similar in form to the orbital decay rate (from $\ket{4}\rightarrow\ket{2}$ and $\ket{3}\rightarrow\ket{1}$) for G4Vs. This corresponds to an effective qubit dephasing rate since the orbital state decay does not affect the spin character whereas for the TLS model it is similar to an amplitude damping channel on the qubit~\cite{Nielsen2010-fr}. The evolution of the G4V electronic states also proceeds similar to the TLS state evolution due to the similarity in the form of the master equation.}

	\section{State Analysis Methods}
	\label{sec:evol_analysis}
	\subsection{Multi-system State Evolution}
	The master equation for the phonon-coupled evolution in Eq.~\eqref{eq:ME_SiV} governs the evolution of a single vacancy center. Specifically, given an initial system state  $ \rho(t=t_1)\equiv \rho_{\mathrm{init.}} $ at $ t=t_1 $, one may interpret the time evolution upto $ t=t_2 $ as a quantum channel acting on the system. In this formalism, we compactly express the state evolution in terms of the Lindbladian for Eq.~\eqref{eq:ME_SiV} as:
	\begin{align}
		\mathcal{L}_{t_2-t_1} \left[\rho_{\mathrm{init.}}\right]=\rho(t_2).
	\end{align}
	For the general treatment of $ N $ such quantum systems, the evolution of the joint spin-state $ \rho^{(N)}_{\mathrm{init.}} $ in the interval $ t\in[t_1,t_2]$ is expressed as,
	\begin{subequations}
		\begin{align}
			\mathcal{L}^{(N)}_{t_2-t_1}[\rho^{(N)}]&:=\sum_{k=1}^N \mathcal{L}^{k}_{t_2-t_1}\left[ \rho^{(N)}_{\mathrm{init.}}\right] \\
			&\equiv \sum_{k=1}^{N} \mathbb{I}^{\otimes{k-1}}\otimes \mathcal{L}_{t_2-t_1}\otimes \mathbb{I}^{\otimes{N-k}} \left[ \rho^{(N)}_{\mathrm{init.}}\right],
		\end{align}
		\label{eq:multi_spin}
	\end{subequations}
	where we use $ \mathcal{L}^k_{t_2-t_1} $ to represent the Lindbladian of the channel acting on the $k$-th system and an identity map $ \mathbb{I} $ on the remaining $ N-1 $ system. In this paper, we primarily focus on the analysis for $N=2$, i.e., a pair of spin qubits in two distinct diamond-vacancy sites that are initialized in some entangled state $ \rho^{(2)}_{\mathrm{init.}} $. Alternatively, we may prescribe a quantum channel for the decoherence with the aid of an operator sum representation of the channel evolution over some specific time. For the evolution of the quantum state over some discrete time interval of length $\Delta t$, we can prescribe a set of Kraus operators as
	\begin{align}
		\begin{split}
			M_0&= \mathbb{I}_{\mathrm{sys}} -\frac{1}{2} \biggl[ \left(\gamma(\bar{n}+1)+\hbar\omega'_A\right)(\outprod{4}+\outprod{3}) \\
			&\qquad \qquad \quad \;+ \left(\gamma(\bar{n})-\hbar\omega'_A\right)(\outprod{2}+\outprod{1}) \biggr]\Delta t,
			\\
			M_1&=\sqrt{\gamma(\bar{n}+1)\Delta t}\,\proj{2}{4},\\
			M_2&=\sqrt{\gamma(\bar{n})\Delta t}\,\proj{4}{2},\\
			M_3&=\sqrt{\gamma(\bar{n}+1)\Delta t}\,\proj{1}{3}, \, {\text{and}}\\
			M_4&=\sqrt{\gamma(\bar{n})\Delta t}\,\proj{3}{1},
		\end{split}
		\label{eq:kraus_ops}
	\end{align}
	with the assumption that terms of order $(\gamma\Delta t)^2$ are negligible; detailed derivation of the same is given in Appendix~\ref{app:kraus_op}. We may correspondingly define a multi-system (composite) Kraus operator set (similar to the definition of $\mathcal{L}^{(N)}_{t_2-t_1}[\cdot]$); we choose the notation $\{\mathbf{M}_{n,k};k=\{0,1,\ldots,4\}\}$, with the definition,
	\begin{align}
		\mathbf{M}_{n,k}=\sum_{l=1}^n \mathbb{I}^{\otimes (l-1)}\otimes M_k\otimes\mathbb{I}^{ \otimes (n-l)},\label{eq:multi_kraus}
	\end{align}
	where each term of the summation applies the $k$-th Kraus operator from Eq.~\eqref{eq:kraus_ops} on the $l$-th defect subsystem.
	
	\subsection{State Quality Evaluation}
	For our analysis of the quantum states under decoherence, we shall use two specific metrics to evaluate the final state quality, namely the state fidelity and the hashing bound. The \emph{state fidelity} $ F(\rho_1,\rho_2) $  between two quantum states $ \rho_1 $ and $ \rho_2 $ evaluates their `overlap' as $F(\rho_1,\rho_2)=\Tr(\sqrt{\sqrt{\rho_1}\rho_2\sqrt{\rho_1}})^2$, which simplifies to $F(\rho_1,\rho_2)=\Braket{\Psi_2|\rho_1|\Psi_2}$ if $\rho_2=\outprod{\Psi_2}$. Fidelity is a reliable and insightful state quality indicator, and generally easy to evaluate. However the `similarity' of states is only applicable in the regime where $ F(\rho_1,\rho_2)\rightarrow 1 $. More task-dependent quantities must be used for state utility analysis, e.g., for quantum communications. 
	
	For the evaluation of bipartite entanglement quality, information theoretic quantities to evaluate (or bound) the distillable entanglement of the state are more insightful. Distillable entanglement, represented by $ E_D(\rho_{AB}) $, quantifies the number of perfect entangled pairs (Bell pairs) that can be distilled from $ \rho_{AB }$, assuming both parties have ideal universal quantum computers (using an arbitrary non-specified distillation circuit) and unlimited two-way classical communications. For general states, $ E_D(\rho_{AB}) $ is non-trivial to evaluate; for the present study we will use the hashing bound $ I(\rho_{AB}) $, which is a lower bound to the state's distillable entanglement and is calculated for the general bipartite state $ \rho_{AB} $ as,
	\begin{align}
		I(\rho_{AB})=\min [S(\rho_A)-S(\rho_{AB}), S(\rho_B)-S(\rho_{AB})],
	\end{align}
	where, $ \rho_A =\Tr _B(\rho_{AB})$, $ \rho_B =\Tr _A(\rho_{AB})$ and $ S(\rho) $ is the von Neumann entropy of the state $ \rho $. 
	
	Analysis of state decoherence shall be focused on two aspects --- (1) the evolution of a single vacancy system whose qubit manifold is  initialized in an arbitrary single qubit state and (2) the evolution of a pair of vacancy centers whose qubit manifold are entangled through a heralded photonic entanglement swap. For the single qubit analysis, we evaluate the quality of a qubit initialized in an equal superposition state. For the latter, we analyze the decoherence of ideal Bell states, as well as realistic models of spin qubits in an entangled pair generated by heralded entangelement swaps~\cite{Dhara2022}. The degradation of the state's hashing bound is our metric of choice for this study.

	
	\section{Spin Decoherence Analysis}
	\label{sec:entanglement_decohere}
	\subsection{Single Spin}
	\label{sec:sing_spin_decoh}
\begin{figure}[htbp]
		\centering
		\includegraphics[width=0.8\linewidth]{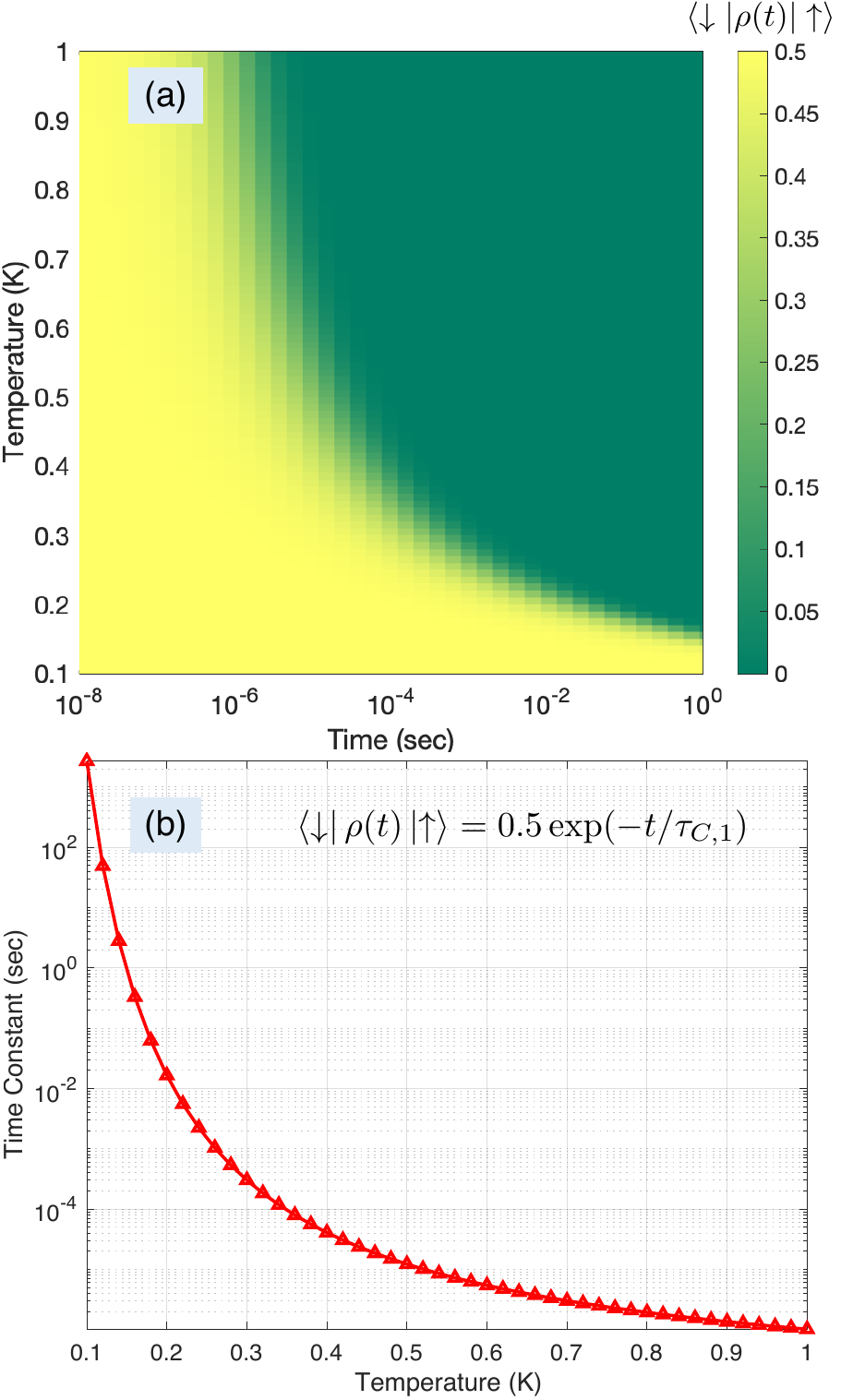}
		\caption{ Evolution of the single spin state coherence   $ \braket{\downarrow\!|\rho(t)|\!\uparrow}= \braket{\uparrow\!|\rho(t)|\!\downarrow} $  initialized in the  $ \ket{\psi(t=0)} =(\ket{1}+\ket{2})/\sqrt{2}$ state. We look at (a) dependence of $ \braket{\downarrow\!|\rho(t)|\!\uparrow} $ at various values of bath temperature $ (T) $ and maximum time, and (b) extract the decay time constant by fitting to the expected relation  $ \braket{\uparrow\!|\rho(t)|\!\downarrow}  = 0.5\times \exp (-t/\tau_{C,1}) $.  We assume $ \Delta= 50 $ GHz (corresponding to Si vacancies in diamond).}
		\label{fig:single_spin_decohere}
	\end{figure}

	For a preliminary understanding of our model, we begin by considering the single spin qubit case. We initialize our spin to the equal superposition state, $ \ket{\psi(0)} =(\ket{1}+\ket{2})/\sqrt{2}$. Evolution for some amount of time $ t $ under the spin-phonon coupled bath model will lead to a mixed state of the spin, which we label by $ \rho(t) $. { We seek to characterize the action of decoherence by studying the decay of off-diagonal term of the electron-spin qubit's density matrix, i.e., either $ \braket{\downarrow\!|\rho(t)|\!\uparrow} $ or $ \braket{\uparrow\!|\rho(t)|\!\downarrow} $, commonly referred to as the \emph{spin coherence} term. The spin degree of freedom is separated from the orbital state by `tracing out' the orbital degree of freedom, i.e., by mapping the states $\{\ket{3},\ket{1}\}\rightarrow\ketdown$ and $ \{\ket{4},\ket{2}\}\rightarrow\ketup$. We expect an exponentially decay of the form $ \braket{\uparrow\!|\rho(t)|\!\downarrow}  = 0.5\times \exp (-t/\tau_{C,1}) $. Henceforth, we shall refer to each of the states by their corresponding spin degree of freedom.}  Fig.~\ref{fig:single_spin_decohere}(a) plots the overall decay of the coherence term for a range of bath temperatures $ T $ for $\Delta=50$ GHz, which governs the mean excitation number of the phonon (bath) environment state $ \rho_{\mathrm{env}} $. We extract $ \tau_{C,1} $ through numerical fitting, in Fig.~\ref{fig:single_spin_decohere}(b) for specified values of $ T $. Indeed we observe an inverse relation between $ T $ and $ \tau_{C,1} $, i.e.\ decoherence times are shorter for higher temperatures, as is expected. The reader may note that $ \tau_{C,1} $ is equally valid for evaluating the fidelity of the state $ \rho(t) $. This is simply because the state fidelity is evaluated as {$ \braket{\psi(0)|\rho(t)|\psi(0)}=0.5+0.5 \exp (-t/\tau_{C,1})$}, since the diagonal terms are unaffected by the phonon interaction. { Additional analysis for heavier G4Vs has been performed in Appendix~\ref{app:heavier_vacancy}.}
	
	\subsection{Ideal Bell Pair}
\begin{figure}[htbp]
		\centering
		\includegraphics[width=0.8\linewidth]{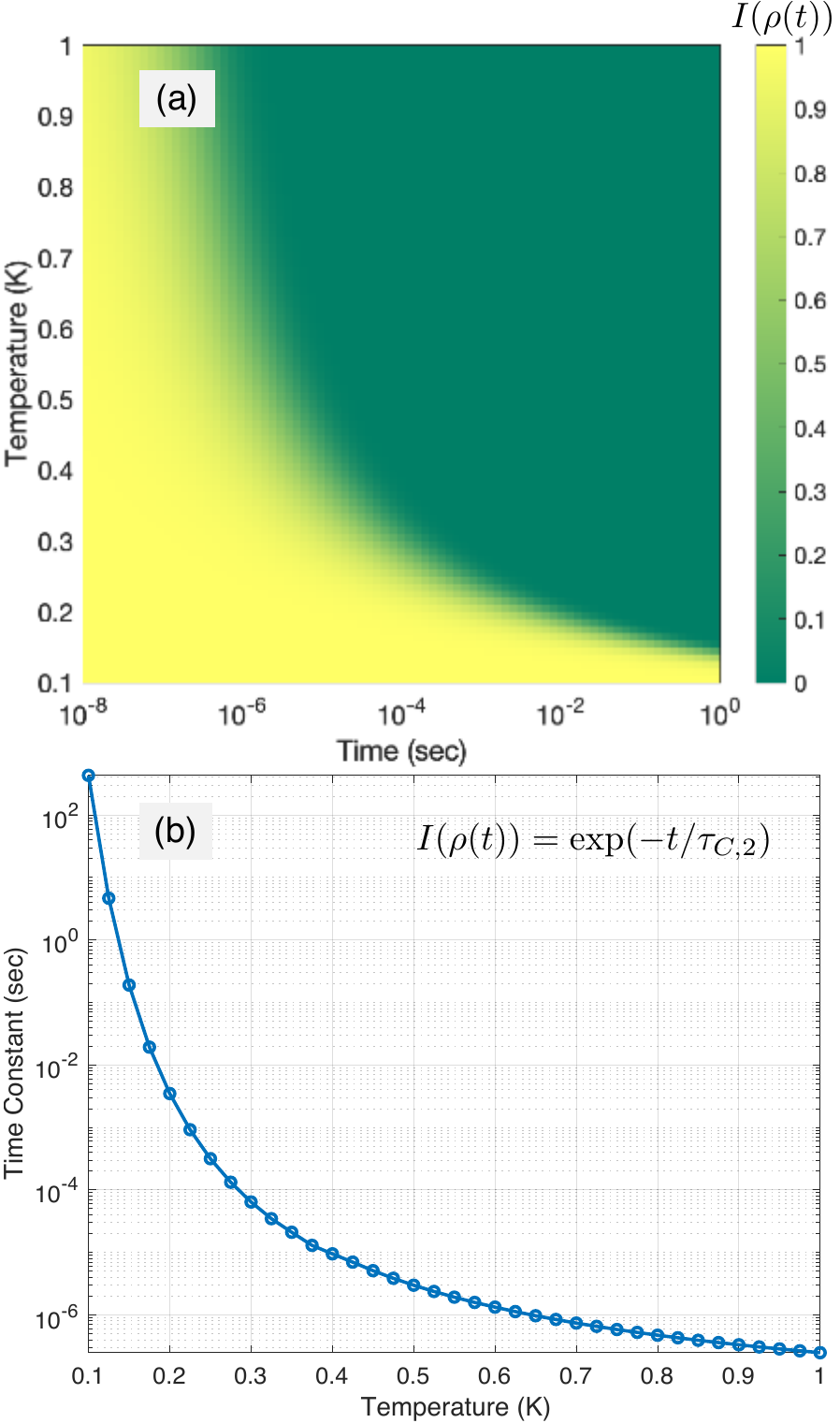}
		\caption{ Evolution of the entangled state hashing bound $ I(\rho(t)) $ for two spins initialized in the  entangled state, $ \ket{\psi^{(2)}(t=0)}= (\ket{1,2}+\ket{2,1})/\sqrt{2}$ . We look at (a) dependence of $ I(\rho(t))$ at various values of bath temperature $ (T) $ and maximum time, and (b) extract the decay time constant  by fitting to the expected relation  $ I(\rho(t)) = \exp (-t/\tau_{C,2}) $.  We assume $ \Delta= 50 $ GHz (corresponding to Si vacancies in diamond).  }
		\label{fig:ent_spin_decohere}
	\end{figure}

	We extend the analysis by looking the evolution of two spins which are initialized in the entangled state $ \ket{\psi^{(2)}(0)}= (\ket{1,2}+\ket{2,1})/\sqrt{2}$. We assume that no time lapses in the initialization of the spin state, and we are able to examine the joint state's decoherence from $ t=0 $. We evaluate the hashing bound  $ I(\rho(t)) $ and proceed similarly to the analysis of the single spin state. Readers should note that the evolution of the joint two-spin state follows the dynamical map formulated in Eqs.~\eqref{eq:multi_spin}.  Fig.~\ref{fig:ent_spin_decohere}(a) plots the overall decay of $ I(\rho(t)) $  for a range of bath temperatures $ T $. Similar to our analysis of the single qubit coherence in Sec.~\ref{sec:sing_spin_decoh}, {we expect an exponential decay of $I(\rho(t))$ with time, i.e.\ $I(\rho(t)) =\exp(-t/\tau_{C,2})$}. We extract $ \tau_{C,2} $ through numerical fitting, in Fig.~\ref{fig:ent_spin_decohere}(b) for specified values of $ T $.  An inverse relation between $ T $ and $ \tau_{C,2} $ is observed, which corroborates the observations we made for the single spin evolution. We note that $ \tau_{C,2}<\tau_{C,1} $ for all $ T $.
	
	\subsection{Distributed Entangled States}
\begin{figure}[ht]
		\centering
		\includegraphics[width=\linewidth]{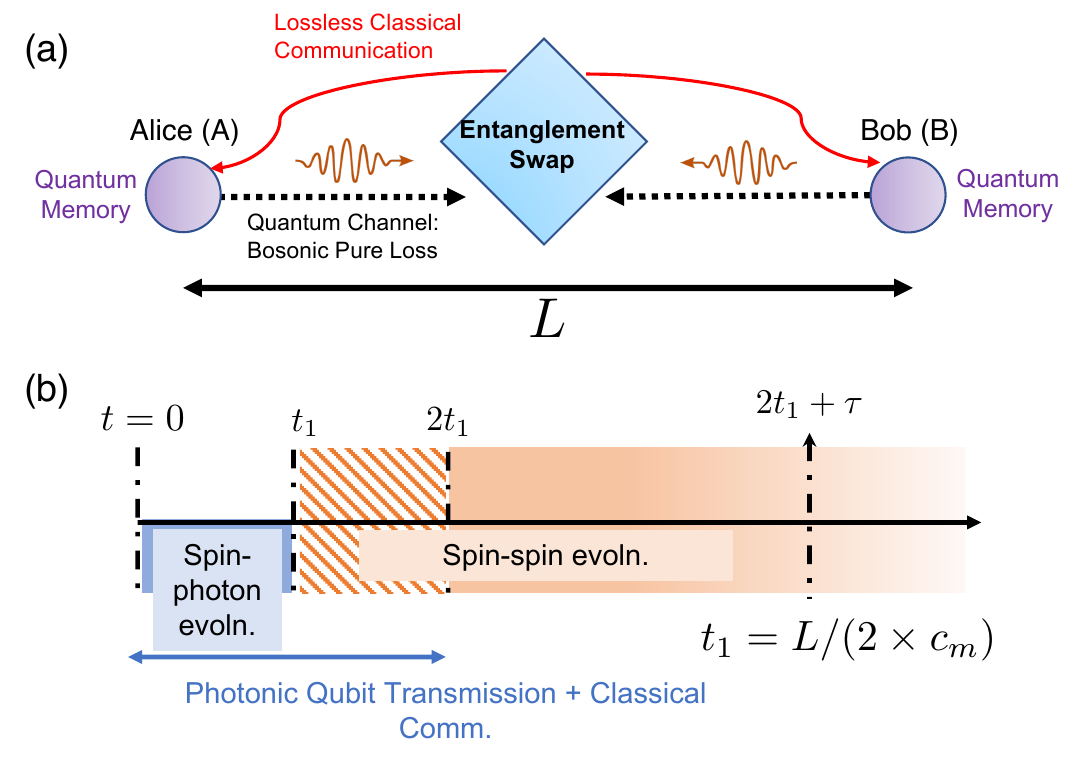}
		\caption{ (a) Layout of `midpoint entanglement swap' between two parties Alice (A) and Bob (B) separated by $ L $ km. We assume a bosonic pure loss channel as the quantum channel over which the qubit is transmitted, additionally we assume a lossless classical communication channel for transmission of entanglement swap heralding information. (b) Timeline of entanglement  generation using  `midpoint entanglement swap' with relevant temporal milestones marked. }
		\label{fig:swap_layout}
	\end{figure}

	Solid state spin based defects are promising candidates for the generation of entanglement over a network. Unlike the generation of local entanglement where the joint state can be used and analyzed from the moment of initialization, accounting for network latency is key for states generated/distributed remotely. This is apparent by considering the simple `midpoint entanglement swap' architecture (depicted in Fig.~\ref{fig:swap_layout}(a)) which is one of two canonical setups for entanglement generation/distribution over a quantum link~\cite{Dhara2022,Wein2020-vp,Hermans2022-je}. 
	As depicted in Fig.~\ref{fig:swap_layout}(a), the midpoint swap link involves two parties Alice (A) and Bob (B), who generate a photonic qubit (brown wavepacket) entangled with their spin qubit (purple circle), which we label as $ \sigma_{\mathrm{spin,photon}} $. The photonic qubit is then transmitted to the `midpoint' (blue diamond) where an entanglement swap takes place. We assume that A and B are separated by a physical network length of $ L $. Assuming that the speed of light in the medium is $ c_m $, and the point of generation of the spin-photonic qubit entanglement is at $ t=0 $, it takes $ t=L/(2c_m) \equiv t_1$ for the photons to travel to the midpoint for the entanglement swap. Hence for $ T\in[0,t_1] $, we account for decoherence of the individual spins using a composite channel on the joint spin-photonic qubit state, say. This channel takes the form 
	\begin{align}
		\sigma_{\mathrm{spin,photon}} {\rightarrow}	 \mathcal{L}_{t_1-0,\mathrm{spin}}\otimes \mathbb{I}_{\mathrm{photon }} [\sigma_{\mathrm{spin,photon}}].
	\end{align}
	At $ t=t_1 $, if the entanglement swapping operation succeeds, the spins are entangled { and} their joint state decoheres as per the joint state evolution rule $ (N=2) $ of Eqs.~\eqref{eq:multi_spin}. However the end users A and B do not immediately have access to this information, since the entanglement swapping measurement outcomes must reach the parties, which takes another $ L/(2 c_m) $ seconds. Thus, any accessible entangled state generated in this form is accessible only after $ t=2t_1=L/c_m $.
	\begin{figure}[htbp]
		\centering
		\includegraphics[width=0.95\linewidth]{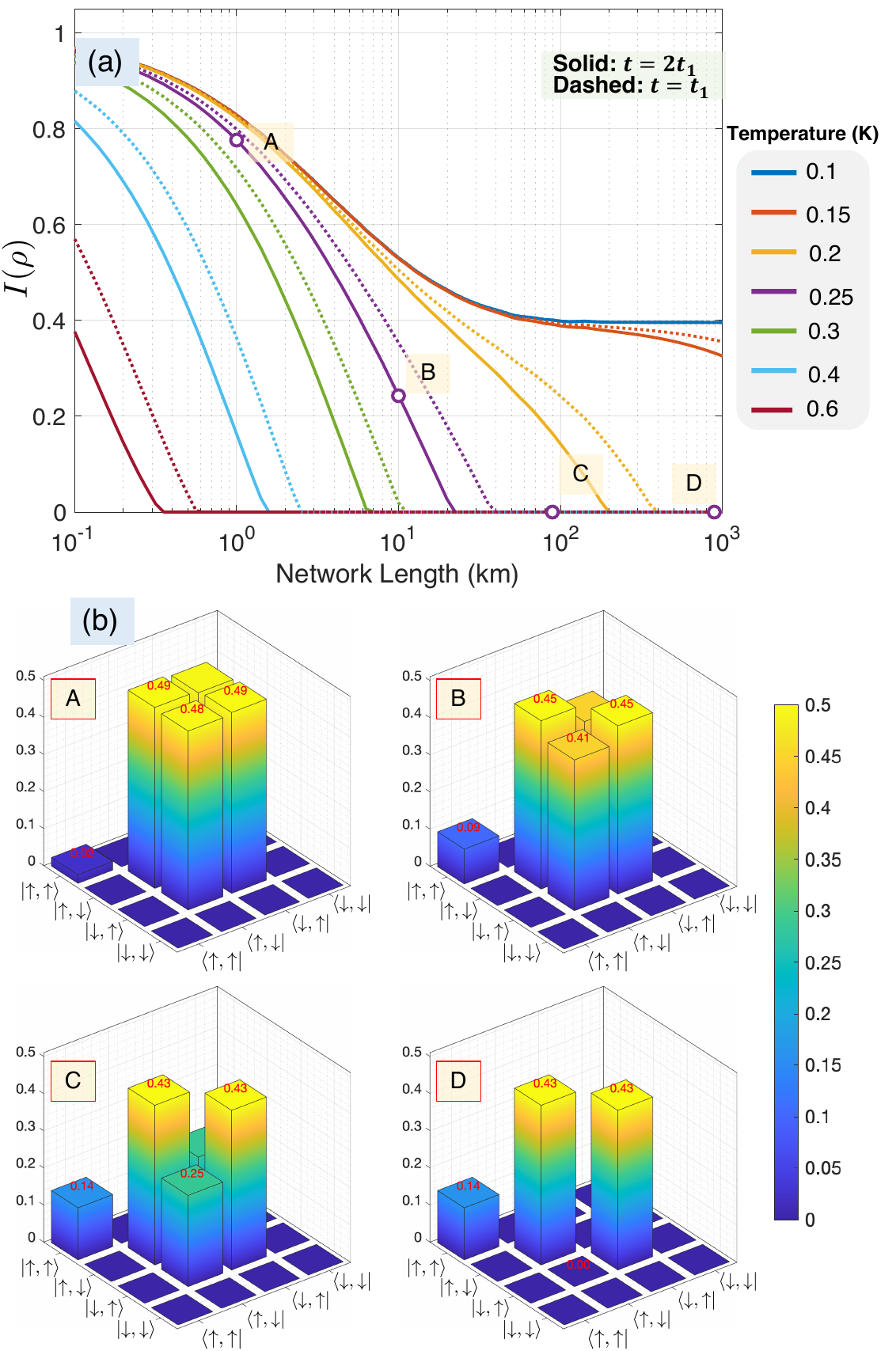}
		\caption{(a) Quality of entangled state generated using the single-rail photonic encoding for `midpoint entanglement swap' quantified  using the hashing bound quantity for states right after entanglement swap ($ t=t_1 $; dashed) and when the heralding information is received by the end users ($ t=2t_1 $; solid) for varying values of temperature. (b)Visualized density matrix of spin-spin entangled state for marked values of $ L $ (A-D) at $ T= 0.250$ K. We assume $ \Delta= 50 $ GHz. }
		\label{fig:singleRail_decohere}
	\end{figure}
	
\begin{figure}[htbp]
		\centering
		\includegraphics[width=0.95\linewidth]{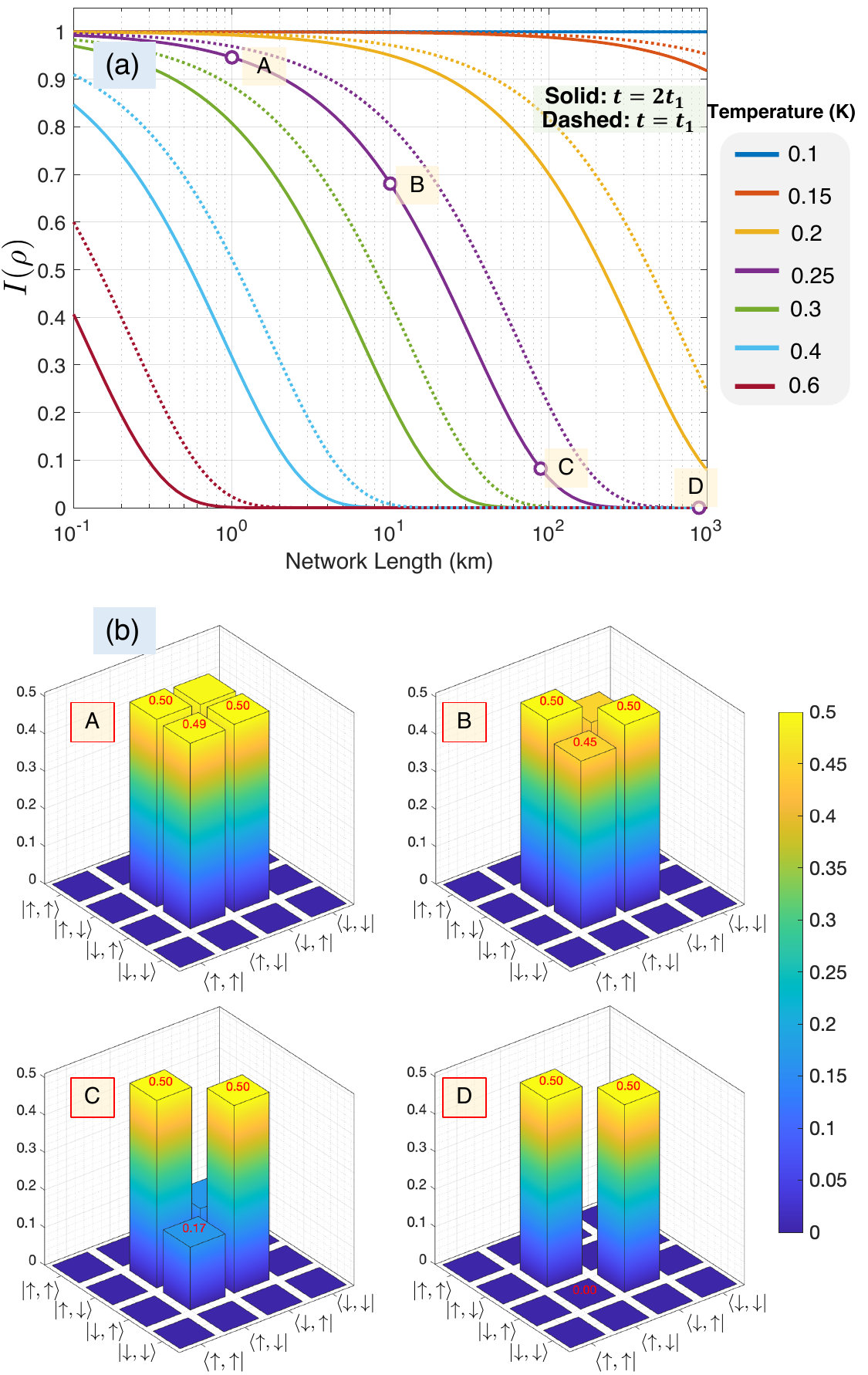}
		\caption{ (a) Quality of entangled state generated using the dual-rail photonic encoding for `midpoint entanglement swap' quantified  using the hashing bound quantity for states right after entanglement swap ($ t=t_1 $; dashed) and when the heralding information is received by the end users ($ t=2t_1 $; solid) for varying values of temperature. (b)Visualized density matrix of spin-spin entangled state for marked values of $ L $  (A-D) at $ T= 0.250$ K. We assume $ \Delta= 50 $ GHz  (corresponding to Si vacancies in diamond).}
		\label{fig:dualRail_decohere}
	\end{figure}

	In Figs.~\ref{fig:singleRail_decohere}-\ref{fig:dualRail_decohere}, we examine the quality of the entangled state under the complete action of decoherence. We use the spin entangled states derived in Ref.~\cite{Dhara2022} and evaluate the state quality by calculating the $ I(\rho) $. We assume that the channel between Alice and Bob is spanned by an optical fiber whose transmissivity scales as $ \eta(L)=\exp(-\alpha L) $ with $ \alpha =13$ dB/km. {The entanglement swapping circuit is noiseless and there is no mode or carrier phase mismatch (i.e., $ P_d=0; \mathcal{V}=1 $ in the formulation of~\cite{Dhara2022}, Sec.~IVA). }
	Subfigures (a) for both Fig.~\ref{fig:singleRail_decohere}-\ref{fig:dualRail_decohere} show the hashing bound of the state when it is heralded (at $ t=2t_1 $; solid) as well as the inaccessible state at the moment of entanglement generation (at $ t=t_1 $; dashed) for varying bath temperatures. For ease of visualization, we choose the specific scenario of $ T=0.250$K to depict bar plots of the final spin density matrices in subfigures (b). The overall effect of decoherence in damping the off-diagonal terms i.e.\ $ \braket{\uparrow,\downarrow\!|\rho|\!\downarrow,\uparrow} $ and $ \braket{\downarrow,\uparrow\!|\rho|\!\uparrow,\downarrow} $ is evident from a visual inspection. Readers may also note that for large $ L $, the value of  $ I(\rho) $ for the single rail heralded case $ (\sim\!0.4 \text{ ebits per copy}) $ is lower than the dual rail heralded state. Comparing corresponding density matrix representations gives us a hint: $ \braket{\uparrow,\uparrow\!|\rho|\!\uparrow,\uparrow}  $ is strictly non-zero for all $ L>0 $ in the  single-rail case. Detailed discussions about the reason behind this clear contrast are given in~\cite{ Dhara2022,Hermans2022-je}.

	{
		
		\subsection{Multipartite Entangled States}
		
		For building fault-tolerant quantum repeaters, as well for distributed quantum computations facilitated by a network, entangled states among multiple spin qubits (i.e., $N>2$ defect centers) will be required. Analyzing the time-dynamics of any application-driven metric of such an $N$-qubit entangled state (as each qubit decoheres due to interaction with their local phonon bath)  using the full Master Equation formalism will require tracking the $4\times4$ -dimensional density matrix for each qubit, which becomes intractable for larger $N$, since the system size scales as $\mathcal{O}(4^{2N})$. Expressing the action of decoherence via a single-qubit channel expressed via the multi-system Kraus operators defined in Eq.~\eqref{eq:multi_kraus}, will significantly simplify such analyses. Below are some example problems that this formalism could be useful for.
		
		
		\begin{enumerate}
			\item Quantifying the time evolution of (bipartite) entanglement across an arbitrary bi-partition of an $N$-qubit entangled state generated via heralded photonic Bell measurements,
			\item Study the differences, if any, between the multi-partite entanglement decay rate for $N$-qubit stabilizer states versus $N$-qubit states that are not stabilizer states (i.e., preparing which requires us to initialize the spins followed by some non-Clifford quantum logic applied on them). 
			\item Quantifying the time threshold when {\em genuine multi-partite entanglement} disappears for an $N$-qubit entangled state,
			\item Quantifying the time evolution of the quantum Fisher information (QFI) of metrologically-useful $N$-spin entangled states, e.g., prepared for entanglement-assisted sensing of a spatially-correlated magnetic field, and
			\item Quantifying the time evolution of an $N$-qubit error correction code (i.e., one that encodes $K < N$ logical qubits) to quantify the time threshold beyond which the code can no longer correct for the collective decoherence-induced error.
		\end{enumerate}
		It is expected that the off-diagonal elements of the quantum state will decay exponentially with a time constant proportional to $N$; however, a rigorous relation to the state's entanglement metric (for e.g. the genuine multipartite entanglement) is not clear at this juncture.
}
	
	\section{Conclusions and Outlook}
	\label{sec:conclusion}
	
	The task of quantifying the effect of decoherence for quantum memories is crucial in understanding their utility in a variety of tasks. Specifically for applications dependent on shared entanglement, the quality of the distribute state is important. Our study on the complete master equation modeling for spin-phonon coupling G4Vs in diamond promotes the necessity to understand the complete quantum state dynamics. We have developed a prescription to track and quantify the state's density operator. Further processing of spin qubits, for tasks such as intra-memory entanglement swap (which are required for repeater networks), entanglement distillation (to boost the quality of the shared entangled states), or distributed quantum computing, would be greatly informed by the complete density operator. Our study is unique in this approach to close the gap between theoretical predictions and various experimental characterizations of various G4Vs.
	
	Phonon coupling is one part of a multitude of decoherence factors relevant to spin qubits based on defect centers in diamond. The effect of the host material's nuclear spin bath in the decoherence of the qubit has been omitted in this study. The vacancy atom's local nuclear spin environment also brings in some non-Markovian characteristics in the evolution of the system. Changes in the experimental setup, for e.g.\, using  off-axis magnetic fields and mechanical strain tuning of spin vacancies also modify the electronic structure of these systems in non-trivial ways. Accounting for these effects in conjunction with phonon coupling in more detailed models will be promising for a variety of applications. We hope that techniques that have been illustrated by our study will motivate such future studies.
	
	\section{Acknowledgments}
	We thank Christos N. Gagatsos (Univ. of Arizona), Kevin C. Chen (MIT; currently at HRL Laboratories), Isaac B.W. Harris (MIT), Hyeongrak Choi (MIT) and Dirk Englund (MIT) for fruitful discussions and comments on the manuscript.
	The authors acknowledge the Mega Qubit Router (MQR) project funded under federal support via a subcontract from the University of Arizona Applied Research Corporation (UA-ARC), for supporting this research. Additionally, the authors  acknowledge National Science Foundation (NSF) Engineering Research Center for Quantum Networks (CQN), awarded under cooperative agreement number 1941583, for synergistic research support. S.G. has outside interests in SensorQ Technologies Incorporated and Guha, LLC. 
	 These interests have been disclosed to UArizona and reviewed in accordance with its conflict of interest policies, with any conflicts of interest to be managed accordingly.
	
	\onecolumngrid
	\appendix 

	\section{Electronic Structure of Group IV Vacancies}
	\label{sec:app_elecstruc}
	
	\begin{figure}[ht!]
		\centering
		\includegraphics[width=0.6\linewidth]{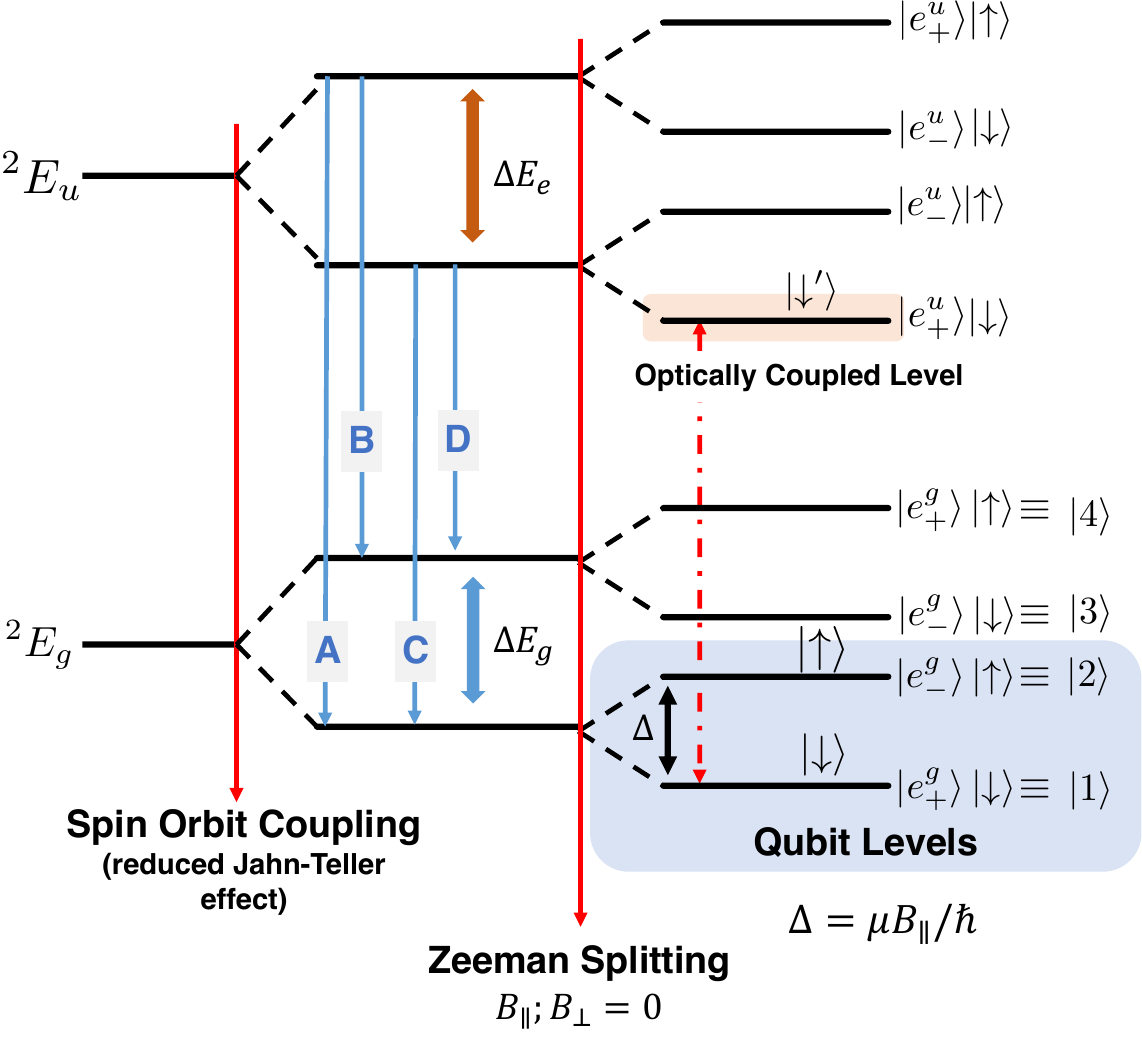}
		\caption{ Electronic level structure of the silicon vacancy center in diamond. The split levels highlighted in blue show the qubit levels.}
		\label{fig:electronic_level}
	\end{figure}
	
	The electronic structure of the group IV vacancies in diamond has been extensively studied in literature~\cite{Hepp2014-fa,Hepp2014-ws,Debroux2021-mw}. The inversion symmetric `staggered ethane' configuration of the vacancy center yields various electronic properties, the most important of which are the isolated energy levels in the diamond bandgap. Interested readers may look at~\cite{Hepp2014-fa,Debroux2021-mw} and the references therein for the detailed analysis of the Si vacancy electronic structure. We start with the complete electronic structure in Fig.~\ref{fig:electronic_level} focus only on the lower branch manifold of the silicon vacancy  (SiV) marked as $ ^{2}E_g $.  $ ^{2}E_g $ is a four-fold degenerate energy level with an orbital degree and a spin degree of freedom. Both quantum mechanical degrees of freedom are two dimensional Hilbert spaces, which allow us to express the complete ground state manifold (alternatively referred to as the lower branch, LB) as the Hilbert space $ \mathcal{H}=\mathcal{H}_{\mathrm{orbital}}\otimes \mathcal{H}_{\mathrm{spin}} $. The logical basis states of $ \mathcal{H}_{\mathrm{orbital}}$ are expressed as $ \ket{e_+} $ and $ \ket{e_-} $, where as for $ \mathcal{H}_{\mathrm{spin}} $, they are $ \ketup $ and $ \ketdown $. However fine structure splitting in the spectra of these systems are not attributable to these bare levels i.e.\ they must account for additional interactions. Theoretical and experimental studies have attributed the splitting in these systems to three major interactions. We discuss subsystem specific interaction terms, and try to frame the problem in terms of the abstracted eigenvectors. Any Hamiltonian described hence forth shall have the generic form,
	\begin{align}
		{H}={H}_{\mathrm{orb.}}\otimes {H}_{\mathrm{spin}}
	\end{align}
	where ${H}_{\mathrm{orb.}}$ and ${H}_{\mathrm{spin}}$ describe the Hamiltonian for the orbital ans spin system respectively.

	\emph{Spin-Orbit Coupling}---The spin-orbit coupling is a relativistic interaction of the electronic orbital with the nuclear potential of the defect atom. Normally a rotation invariant interaction, the crystal field of the host diamond breaks the symmetry for group IV vacancies to yield an interaction that affect orbital eigenstates with energy shifts (without mixing) of the spin-levels. The spin orbit coupling Hamiltonian $ H_{SO} $ is given as,
	\begin{subequations}
		\begin{align}
			{H}_{\mathrm{SO}}&=-\frac{\lambda_{\mathrm{SO}}}{2} (-\proj{e_+}{e_+}+ \proj{e_-}{e_-})\otimes (\outprod{\uparrow}-\outprod{\downarrow}) \\
			&=+\frac{\lambda_{\mathrm{SO}}}{2} \cdot  \hat{Z}_{\mathrm{orb.}}\otimes \hat{Z}_{\mathrm{spin}},
		\end{align}
	\end{subequations}
	where $ \lambda_{\mathrm{SO}} $ is the spin-orbit coupling strength with $ \hat{Z}_{\mathrm{orb.}} = (\proj{e_+}{e_+}-\proj{e_-}{e_-}) $ and $ \hat{Z}_{\mathrm{spin}}= \outprod{\uparrow}-\outprod{\downarrow} $. Hence the joint eigenstates for $ H_{\mathrm{SO}} $ are given as,
	\begin{subequations}
		\begin{align}
			&\{\ket{e_+} \otimes \ketup,  \ket{e_-}\otimes \ketdown\}: \text{with eigenvalue }\lambda_{\mathrm{SO}}/2\\
			&\{\ket{e_+}\otimes \ketdown, \ket{e_-}\otimes \ketup\}: \text{with eigenvalue }-\lambda_{\mathrm{SO}}/2
		\end{align}
	\end{subequations}
	
	\emph{Jahn-Teller interaction} --- This effect introduces distortion of the electronic orbitals due to an asymmetric potential, leading to orbital energy shifts less prominent than spin-orbit coupling. This is a spin-independent interaction with the Hamiltonian
	\begin{subequations}
		\begin{align}
			H_{\mathrm{JT}}&=\left[\Upsilon_x \hat{Y}_{\mathrm{orb.}} -\Upsilon_y \hat{X}_{\mathrm{orb.}}\right] \otimes \hat{\mathbb{I}}_{\mathrm{spin}},
		\end{align}
	\end{subequations}
	where $ \Upsilon_x,\Upsilon_y $ represent the effective energies associated with the distorted potential along the $ x,y $ directions (in the cardinal frame) respectively, and $\hat{Y}_\mathrm{orb.}=i\proj{e_+}{e_-}-i\proj{e_-}{e_+},\hat{X}_\mathrm{orb.}=\proj{e_+}{e_-}+\proj{e_-}{e_+} $. The directions are chosen with respect to the $ \langle111\rangle $ axis of the defect center i.e. the $ z $- axis is along this direction and the rest are chosen according a standard right-handed rotation rule.

	\emph{Zeeman Splitting}--- Zeeman splitting is observed when external magnetic  fields lifts spin-degeneracy of the system. There are two distinct effects dependent on the direction of the field, which may be parallel $ (B_\parallel) $ or perpendicular $ (B_\perp) $ to the high symmetry axis ($ [1,1,1] $ direction) of the defect center. The parallel field yields an effective Hamiltonian, 
	\begin{align}
		H_{Z,\parallel}=\frac{\gamma_e}{2}\cdot \hat{\mathbb{I}}_{\mathrm{orb.}} \otimes B_\parallel \hat{Z}_{\mathrm{spin}}.
	\end{align}
	The perpendicular field yields the effective Hamiltonian,
	\begin{align}
		H_{Z,\perp}=\frac{\gamma_e}{2}\cdot \hat{\mathbb{I}}_{\mathrm{orb.}} \otimes\left[ B_x \hat{X}_{\mathrm{spin}} + B_y\hat{Y}_{\mathrm{spin}}\right] , 
	\end{align}
	Here, $ \gamma_e=2\mu_B /\hbar $ and $ B_x , B_y$ are the orthogonal components of perpendicular field ($ B_\perp $).
	
	The parallel field $ B_\parallel $ does not cause any spin state mixing and only induces a spin-dependent energy shift of $ \pm \gamma_e B_\parallel/2 $. However, a perpendicular magnetic field can cause some spin mixing. The eigenstates of the joint system when we consider $ H_{\mathrm{SO}}+H_{Z,\parallel}+H_{Z,\perp} $ are given by 
	\begin{subequations}
		\begin{align}
			&\ket{e_+}\otimes \left[\ketup +\frac{\gamma_e B_+}{\gamma_e B_z+\lambda_{\mathrm{SO}}+ \sqrt{|\gamma_e B_+|^2 +(\lambda_{\mathrm{SO}}+\gamma_e B_z)^2}}\ketdown \right]\\
			&\ket{e_+}\otimes \left[\ketdown+\frac{\gamma_e B_+}{\gamma_e B_z+\lambda_{\mathrm{SO}}+ \sqrt{|\gamma_e B_+|^2 +(\lambda_{\mathrm{SO}}+\gamma_e B_z)^2}}\ketup\right]\\
			&\ket{e_-}\otimes \left[\ketup -\frac{\gamma_e B_+}{-\gamma_e B_z+\lambda_{\mathrm{SO}}+ \sqrt{|\gamma_e B_+|^2 +(\lambda_{\mathrm{SO}}+\gamma_e B_z)^2}}\ketdown \right]\\
			&\ket{e_-}\otimes \left[\ketdown-\frac{\gamma_e B_+}{-\gamma_e B_z+\lambda_{\mathrm{SO}}+ \sqrt{|\gamma_e B_+|^2 +(\lambda_{\mathrm{SO}}+\gamma_e B_z)^2}}\ketup\right]
		\end{align}
	\end{subequations}
	where $B_z=B_\parallel;  B_+=B_x+iB_y $. Hence the total field is $ B=\sqrt{B_x^2+B_y^2+B_z^2} $. For the present article, we assume that $ B_+=0 $ i.e. the Zeeman levels are not spin mixed. We then choose the following abbreviations for the states (based on  their energy levels) as
	
	\begin{align}
		\begin{split}
			\ket{1}\equiv\ket{e_+}\otimes\ketdown;\quad \ket{2}\equiv\ket{e_-}\otimes\ketup\\
			\ket{3}\equiv\ket{e_-}\otimes\ketdown;\quad \ket{4}\equiv\ket{e_+}\otimes\ketup
		\end{split}	
	\end{align}
	
	\section{Derivation of Phonon-Spin Coupling Master Equation }
	\label{app:deriv_ME}
	\subsection{Background}
	\label{app:meq_background}
	We shall (without detailed discussion) describe the master equation derivation~\cite{Breuer2002-op,Carmichael1999-zf}. We consider (1) the system of interest (labelled by the subscript $ \sys$) with a density operator $ \rho_{\sys} $ evolving under the Hamiltonian $ H_{\sys} $; (2) an environment system (labelled by the subscript $ \env $) with a density operator $ \rho_\env$ evolving under the Hamiltonian $ H_\env $; (3) an interaction between the system and environment governed by the interaction Hamiltonian $ H_{\itn} $. Hence, the total Hamiltonian is of the form $ H=H_\sys+H_\env+ H_\itn $.
	
	One may look at the evolution of the system+reservoir density operator $ \rho_{\sys-\env} $ (which, in general, may not be factorizable) under the interaction by using the interaction picture definitions, 
	\begin{align}
		&\tilde{\rho}_{\sys-\env}(t) \equiv e^{(i / \hbar)\left(H_{\sys}+H_{\env}\right) t} \rho_{\sys-\env}(t) e^{-(i / \hbar)\left(H_{\sys}+H_{\env}\right) t}\\
		&\tilde{H}_{\sys-\env}(t) \equiv e^{(i / \hbar)\left(H_{\sys}+H_{\env}\right) t} H_{\sys-\env} e^{-(i / \hbar)\left(H_{\sys}+H_{\env} \right) t}
	\end{align}
	Then the exact evolu tion of the joint system is given by the master equation
	\begin{align}
		\dot{\tilde{\rho}}_{\sys-\env}=\frac{1}{i \hbar}\left[\tilde{H}_{\sys-\env}(t), \rho_{\sys-\env}(0)\right]-\frac{1}{\hbar^{2}} \int_{0}^{t} d t^{\prime}\left[\tilde{H}_{\sys-\env}(t),\left[\tilde{H}_{\sys-\env}(t'), \tilde{\rho}_{\sys-\env} (t')\right]\right],
		\label{eq:master_eqn}
	\end{align}
	which under assumptions of initial state separability and Born-Markov approximations, becomes the master equation
	\begin{align}
		\dot{\tilde{\rho}}_{\sys}=-\frac{1}{\hbar^{2}} \int_{0}^{t} d t' \Tr_R\left\{\left[\tilde{H}_{\sys-\env}(t),\left[\tilde{H}_{\sys-\env}(t'), \tilde{\rho}_{\sys} (t) \otimes \rho_\env \right]\right] \right\}.
	\end{align}
	\label{eq:BM_master}
	
	Further we use the system-environment operator decomposition of the master equation~\cite{Carmichael1999-zf}. Specifically, if the $ H_{\itn} $ can be written as
	\begin{align}
		&H_{\sys-\env}=\hbar \sum_i s_i \Gamma_i \quad \text{(Schr\"{o}dinger Picture)}\\
		& \Leftrightarrow \tilde{H}_{\sys-\env} (t)=\hbar \sum_i \tilde{s}_i (t) \tilde{\Gamma}_i (t) \quad\text{(Interaction Picture)}
		\label{eq:sys_res_int}
	\end{align}
	for some system operators $ s_i $ and environment operators $ \Gamma_i $, 
	\begin{align}
		&\tilde{s}_i(t)= e^{(i / \hbar) H_{\sys} t}\; s_{i}\; e^{-(i / \hbar) H_{\sys} t}\\
		&\tilde{\Gamma}_i(t)= e^{(i / \hbar) H_{\env} t} \; \Gamma_{i}\; e^{-(i / \hbar) H_{\env} t}
	\end{align} 	
	then the master equation for the system density operator $ \rho_S $ can be written as,
	\begin{align}
		\begin{split}
			\dot{\tilde{\rho}}_\sys
			=-\sum_{i, j} \int_{0}^{t} d t^{\prime}\biggl\{& \left[\tilde{s}_{i}(t) \tilde{s}_{j}(t') \tilde{\rho}_\sys (t')-\tilde{s}_{j}(t') \tilde{\rho}_\sys (t') \tilde{s}_{i}(t)\right]\left\langle\tilde{\Gamma}_{i}(t) \tilde{\Gamma}_{j}(t')\right\rangle_{\env}\\
			&+\left[\tilde{\rho}_\sys (t') \tilde{s}_{j}(t') \tilde{s}_{i}(t)-\tilde{s}_{i}(t) \tilde{\rho}_\sys (t') \tilde{s}_{j}(t')\right]\left\langle\tilde{\Gamma}_{j}(t') \tilde{\Gamma}_{i}(t)\right\rangle_{\env}\biggr\}.
		\end{split}
		\label{eq:simplified_BM_master}
	\end{align}
	
	\subsection{Spin Decoherence: Formulation and Solution}
	\label{app:meq_formulation}
	\begin{figure}[ht!]
		\centering
		\includegraphics[width=0.5\linewidth]{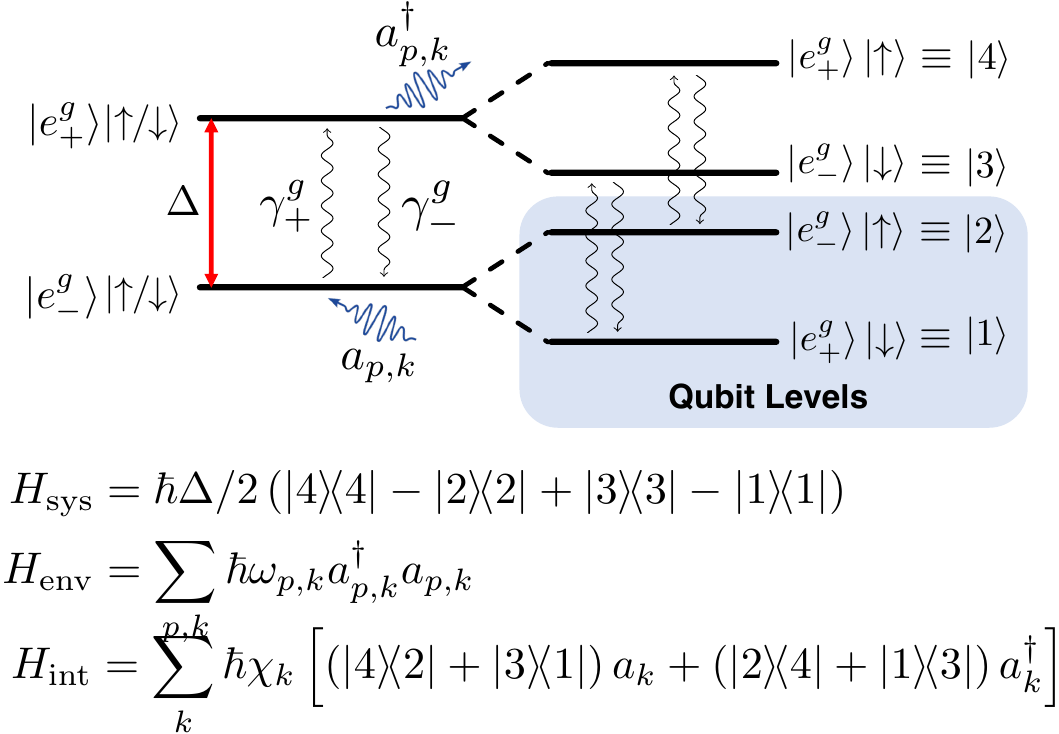}
		\caption{ Phonon coupling with split levels of the ground state manifold of the G4V center.}
		\label{fig:phonon_coupling_app}
	\end{figure}
	
	For the phonon coupling problem, we may identify four pairs of system reservoir operators (in line with the formulation of Eq.~\eqref{eq:sys_res_int}) as  
	\begin{subequations}
		\begin{align}
			&s_1=\proj{4}{2}; \quad\Gamma_1=\sum_k \chi_k \hat{a}_{k}\\
			&s_2=\proj{3}{1}; \quad \Gamma_2= \sum_k \chi_k \hat{a}_{k}\\
			&s_3=\proj{2}{4}; \quad \Gamma_3=\sum_k \chi_k \hat{a}^{\dagger}_{k}\\
			&s_4=\proj{1}{3}; \quad \Gamma_4=\sum_k \chi_k \hat{a}^{\dagger}_{k}.
		\end{align}
	\end{subequations}
	We note that $ s_3=s_1^\dagger; s_4=s_2^\dagger $. Identifying that $\Gamma_1=\Gamma_2; \Gamma_3=\Gamma_4 $ and $ \Gamma_3=\Gamma_1^\dagger $. Hence let us use $ \Gamma=\Gamma_1 $ and correspondingly $ \Gamma^\dagger=\Gamma_3 $.  
	The full interaction picture Hamiltonian then becomes 
	\begin{align}
		\tilde{H}_{\mathrm{sys-env.}} (t)=\sum_{k} \hbar \chi_{k}\left[\left(\proj{4}{2}+\proj{3}{1}\right)a_{ k} e^{i(\Delta-\omega_k)t}+\left(\proj{2}{4}+\proj{1}{3}\right) e^{-i(\Delta-\omega_k)t} a_{k}^{\dagger}\right]
	\end{align}
	
	Examining Eq.~\eqref{eq:simplified_BM_master}, we have obtained the system operators $ \tilde{s}_i$. The next step is to evaluate the reservoir correlation integrals,
	\begin{subequations}
		\begin{align}
			\left\langle\tilde{\Gamma}^{\dagger}(t) \tilde{\Gamma}^{\dagger}(t')\right\rangle_{\env} &=
			\left\langle\tilde{\Gamma}(t) \tilde{\Gamma}(t')\right\rangle_{\env} =0 \\
			\left\langle\tilde{\Gamma}^{\dagger}(t) \tilde{\Gamma}(t')\right\rangle_{\env} &=\sum_{j, k} \chi_{j} \chi_{k} e^{i \omega_{j} t} e^{-i \omega_{k} t^{\prime}}  \Tr_{\env}\left({\rho_\env}  a_{j}^{\dagger} a_{k}\right) \nonumber\\
			&=\sum_{j}\left|\chi_{j}\right|^{2} e^{i \omega_j \left(t-t^{\prime}\right)} \bar{n}\left(\omega_j, T\right) \\
			\left\langle\tilde{\Gamma}(t) \tilde{\Gamma}^{\dagger}(t')\right\rangle_{\env} &=\sum_{j, k} \chi_{j} \chi_{k}^{*} e^{-i \omega_{j} t} e^{i \omega_{k} t^{\prime}} \Tr_{\env}\left({\rho_\env} a_{j} a_{k}^{\dagger}\right) \nonumber\\
			&=\sum_{j}\left|\chi_{j}\right|^{2} e^{-i \omega_j \left(t-t^{\prime}\right)}\left[\bar{n}\left(\omega_j, T\right)+1\right].
		\end{align}
	\end{subequations}
	where $ \bar{n}(\omega_j,T)=  {e^{-\hbar \omega_j / k_{B} T}}/({1-e^{-\hbar \omega_j/ k_{B} T}})$ i.e. the mode occupation follows the Bose-Einstein distribution. Let us explicitly start writing the terms out for our operators. Let us consider $ i=1 $, and following the reservoir correlation functions obtained previously, $ j=3,4 $ yield non-zero correlations.  Considering $ j=3 $,
	
	\begin{align}
		&\left[\tilde{s}_1 (t) \tilde{s}_3 (t') \tilde{\rho} (t') -  \tilde{s}_3 (t') \tilde{\rho} (t') \tilde{s}_1 (t) \right] \langle \Gamma(t) \Gamma^\dagger(t') \rangle_\env + 
		\left[ \tilde{\rho} (t') \tilde{s}_3 (t') \tilde{s}_1 (t) -  \tilde{s}_1 (t) \tilde{\rho} (t') \tilde{s}_3 (t') \right] \langle  \tilde{\Gamma}^\dagger(t') \tilde{\Gamma}(t) \rangle_\env \nonumber\\
		&=\sum_k |\chi_k|^2 e^{-i\omega_k(t-t')}\biggl\{ \left(\bar{n}\left(\omega_k, T\right)+1\right)	\left[\tilde{s}_1 (t) \tilde{s}_3 (t') \tilde{\rho} (t') -  \tilde{s}_3 (t') \tilde{\rho} (t') \tilde{s}_1 (t) \right] + \bar{n}\left(\omega_k, T\right)
		\left[ \tilde{\rho} (t') \tilde{s}_3 (t') \tilde{s}_1 (t) -  \tilde{s}_1 (t) \tilde{\rho} (t') \tilde{s}_3 (t') \right] \biggr\} \label{eq:simp_step1}
	\end{align}
	
	Note that this is still of the Born form i.e. the equation is still not \emph{memoryless}, since Eq.~\eqref{eq:simp_step1} has terms of the form $ \rho(t') $ which depend time up to $ t $. The standard  procedure after this is to make the change of variable $ t'\rightarrow t-\tau $. The summation over modes $ \Sigma_k(\cdot) $ can be replaced by a density of modes integral of the form $ \Sigma_k \kappa(\omega_k)=\int d \omega \kappa(\omega) g(\omega) $ where $ g(\omega) d\omega$ is the number of modes in the interval $ (\omega,\omega+d\omega) $. By choosing the appropriate scaling for $ |\chi(\omega) |^2=\chi_0 \omega$ and $g(\omega)=g_0 \omega^2$~\cite{Jahnke2015-no}, and various insights drawn about the integral of the correlation functions ~\cite{Carmichael1999-zf}, one can make simplifications by identifying that we account for a pair of \emph{radiatively damped two-level system}. The two separate two-level systems are $ \{\ket{2},\ket{4} \}$ and $ \{\ket{1},\ket{3}\} $ and their master equation is then given by ---
	\begin{align}
		\begin{split}
			\dot{\rho}=&-i \frac{1}{2} \omega_{A}^{\prime}\left[(\outprod{4}-\outprod{2}+\outprod{3}-\outprod{1}), \rho\right]\\
			&+\frac{\gamma}{2}(\bar{n}+1)\left(2\times  \ket{2} \!\!\braket{4|\rho|4} \!\! \bra{2} -\outprod{4}\rho-\rho \outprod{4}\right) 
			+\frac{\gamma}{2} \bar{n}\left(2 \times \ket{4} \!\!\braket{2|\rho|2} \!\! \bra{4}-\outprod{2} \rho-\rho \outprod{2} \right)\\
			&+\frac{\gamma}{2}(\bar{n}+1)\left(2\times  \ket{1} \!\!\braket{3|\rho|3} \!\! \bra{1} -\outprod{3}\rho-\rho \outprod{3}\right) 
			+\frac{\gamma}{2} \bar{n}\left(2 \times \ket{3} \!\!\braket{1|\rho|1} \!\! \bra{3}-\outprod{1} \rho-\rho \outprod{1} \right)
			\label{eq:master_eq_v1}
		\end{split}
	\end{align}
	where $  \bar{n}=\bar{n}(\Delta,T)= {e^{-\hbar \Delta / k_{B} T}}/({1-e^{-\hbar \Delta / k_{B} T}})$ and $ \gamma =2\pi g(\Delta)\cdot|\chi (\Delta)|^2=2\pi g_0 \chi_0 \Delta^3 $. The energy separation $\Delta$ is modified by a specific amount $ \omega'_A=\Delta+2\Delta'+\Delta_{\mathrm{Lamb}}  $, where the $ \Delta' $ is a 
	temperature-dependent shift and $ \Delta_{\mathrm{Lamb}} $ is the normal Lamb shift.
	These modifications to the energy splitting arises from quantum vacuum fluctuations and manifests in the environment correlation integrals~\cite{Carmichael1999-zf}.

	\subsection{ Reframing the Master Equation in the Fock-Liouville Notation}
	\label{app:meq_FockLiouville}
	Given the finite-dimensionality of the system, it is easy to convert the master equation in Eq.~\eqref{eq:master_eq_v1} to a set of differential equations. Specifically, we look at the the density operator derivative element wise by evaluating $ \dot{\rho}_{i,j}  =\braket{i|\dot{\rho}|j} $. We obtain the following differential equations by a `brute force' evaluation (Ref.~\cite{Carmichael1999-zf} Ch.~2.2.3 for standard two level solutions), 
	\begin{subequations}
		\begin{align}
			\begin{split}
				\text{For the two-level system formed by levels $\ket{1}$ and $\ket{3}$: }&
				\dot{\rho}_{11}= -\gamma \bar{n} \rho_{11} + \gamma(\bar{n}+1)\rho_{33} \\
				& \dot{\rho}_{33}= +\gamma \bar{n} \rho_{11} - \gamma(\bar{n}+1)\rho_{33} \\
				& \dot{\rho}_{13} = -\left[ \gamma(2\bar{n}+1)/2 - i\omega_A' \right] \rho_{13};\\
				& \dot{\rho}_{31}=\dot{\rho}^\dagger_{13}
			\end{split}\\
			\begin{split}
				\text{For the two-level system formed by level $\ket{2}$ and $\ket{4}$: }&
				\dot{\rho}_{22}= -\gamma \bar{n} \rho_{22} + \gamma(\bar{n}+1)\rho_{44} \\
				& \dot{\rho}_{44}= +\gamma \bar{n} \rho_{22} - \gamma(\bar{n}+1)\rho_{44} \\
				& \dot{\rho}_{24} = -\left[ \gamma(2\bar{n}+1)/2 - i\omega_A' \right] \rho_{24};\\
				& \dot{\rho}_{42}=\dot{\rho}^\dagger_{24}
			\end{split}\\
			\begin{split}
				\text{Other cross terms: }&
				\dot{\rho}_{12}= -\gamma \bar{n} \rho_{12} \\
				& \dot{\rho}_{34}= -\gamma (\bar{n}+1) \rho_{34}  \\
				&\dot{\rho}_{14}= -\left[ \gamma(2\bar{n}+1)/2 - i\omega_A' \right] \rho_{14}\\
				& \dot{\rho}_{23} = -\left[ \gamma(2\bar{n}+1)/2 - i\omega_A' \right] \rho_{23}
			\end{split}
		\end{align}
	\end{subequations}
	The equations are a set of coupled differential equations that may be solved generally by a \emph{coupled eigenvector method}. We write the set of differential equations for the density matrix elements (say of generalized dimension $ n\times n $) in a vectorized (column-vector) format as 
	\begin{align}
		\left[\rho_{i,j}\right]_{n\times n} \mapsto [\rho_{1,1}, \rho_{1,2},\cdots, \rho_{n,n}]^{T} \equiv \Sket{\rho}.
	\end{align}
	{ This is also called the Fock-Liouville notation for the density operator~\cite{Manzano2020-bi,Gyamfi2020-pl}. }The coupled set of first order ODEs can now be expressed as $ \partial_t{\Sket{\rho}}=\tilde{\mathcal{L}} \Sket{\rho} $, where $\tilde{\mathcal{L}} $ is commonly known as the { Liouvillian superoperator matrix~\cite{Manzano2020-bi,Gyamfi2020-pl}}.The most general $\tilde{\mathcal{L}}$ is complex, non Hermitian, and non-symmetric.
	
	If $ \tilde{\mathcal{L}} $ is non singular, we may solve for its eigenvectors $ \{\Sket{e_i}\}; i\in \{1,2,\ldots, n^2\} $ and corresponding eigenvalues $ \{\lambda_i\} $. The initial state of the system $ \Sket{\rho(t=0)} $ when expressed in this eigenbasis is,
	\begin{align}
		\rho(t=0)\mapsto\Sket{\rho(t=0)} = \sum_{k=1}^{n^2} \alpha_k \Sket{{e}_k},
	\end{align}
	where the coefficients $ \{\alpha_i\} $ are determined by the matrix equation, 
	\begin{align}
		\left[\alpha_i\right]_{n^2\times 1} =\left[\Sket{e_1} \, , \Sket{e_2} ,\cdots,\Sket{e_{n^2}}\right]^{-1}\cdot \vec{\rho}_{(t=0)} 
	\end{align} 
	and the general solution of the coupled ODEs is given as
	\begin{align}
		\Sket{\rho(t)}=\sum_{k=1}^{n^2} \alpha_k \Sket{e_k} \exp(-\lambda_k t) .
	\end{align}
	
	For all our analysis, we use the \emph{row-major order} of vectorizing our density matrix, i.e.\,
	$  \left[\rho_{i,j}\right]_{n\times n}  $ becomes
	\begin{align}
		\begin{pmatrix}
			\rho_{1,1} & \rho_{1,2} &  \dots  & \rho_{1,n} \\
			\rho_{2,1} & \rho_{2,2} & \dots  & \rho_{2,n} \\
			\vdots & \vdots & \ddots & \vdots \\
			\rho_{n,1} & \rho_{n,2} & \dots  & \rho_{n,n}
		\end{pmatrix}
		\longmapsto [	\rho_{1,1} ,\,\rho_{1,2},\,  \dots,\,  \rho_{1,n},\,\rho_{2,1},\,\dots,\rho_{n,n}]^T
	\end{align}
	
	\emph{Single spin decoherence} map matrix (say $ \tilde{\mathcal{L}}^{(s)} $) elements in the $ (i,j) $-th are given as,
	\begin{subequations}
		\begin{align}
			(1,1)=(2,2)=(5,5)=(6,6)=-(11,1)=-(16,6)\equiv&-\gamma_+ (\Delta,T)\\
			(11,11)=(12,12)=(15,15)=(16,16)=-(1,11)=-(6,16)\equiv&-\gamma_-(\Delta,T)\\
			(3,3)=(4,4)=(7,7)=(8,8)\equiv &-\gamma_+ (\Delta,T)-\gamma_- (\Delta,T)+2\pi i\Delta\\
			(9,9)=(10,10)=(13,13)=(14,14)\equiv &-\gamma_+ (\Delta,T)-\gamma_- (\Delta,T)-2\pi i\Delta
		\end{align}
	\end{subequations}
	
	\section{Kraus Operator Representation of Spin Decoherence}
	\label{app:kraus_op}
	{ We begin by considering the Born-Markov master equation of Eq.~\eqref{eq:master_eq_v1} as derived in Appendix~\ref{app:meq_formulation}
		\begin{align}
			\begin{split}
				\dot{\rho}=&-i \frac{1}{2} \omega_{A}^{\prime}\left[(\outprod{4}-\outprod{2}+\outprod{3}-\outprod{1}), \rho\right]\\
				&+\frac{\gamma}{2}(\bar{n}+1)\left(2\times  \ket{2} \!\!\braket{4|\rho|4} \!\! \bra{2} -\outprod{4}\rho-\rho \outprod{4}\right) 
				+\frac{\gamma}{2} \bar{n}\left(2 \times \ket{4} \!\!\braket{2|\rho|2} \!\! \bra{4}-\outprod{2} \rho-\rho \outprod{2} \right)\\
				&+\frac{\gamma}{2}(\bar{n}+1)\left(2\times  \ket{1} \!\!\braket{3|\rho|3} \!\! \bra{1} -\outprod{3}\rho-\rho \outprod{3}\right) 
				+\frac{\gamma}{2} \bar{n}\left(2 \times \ket{3} \!\!\braket{1|\rho|1} \!\! \bra{3}-\outprod{1} \rho-\rho \outprod{1} \right)
				\label{eq:master_Lindblad}
			\end{split}
		\end{align}
		Comparing to the standard Lindblad (or Gorini-Kossakowski-Sudarshan-Lindblad; GKSL) formulation~\cite{Gorini1976-ss,Lindblad1976-lr} of a master equation
		\begin{align}
			\dot{\rho}=-\frac{i}{\hbar} [H_{\mathrm{sys}},\rho] +\sum_i\gamma_i \left(L_i \rho L_i^\dagger-\frac{1}{2}\left\{ L_i^\dagger L_i,\rho\right\} \right)
		\end{align}
		we identify the (modified) system Hamiltonian, $H_{\mathrm{sys}}=\hbar\omega'_A/2(\outprod{4}-\outprod{2}+\outprod{3}-\outprod{1})$, along with the Lindblad operators ($\{L_i\}$) and their corresponding rates ($\{\gamma_i\}$)
		\begin{subequations}
			\begin{align}
				L_1=\proj{2}{4};&\quad \gamma_1=\gamma(\bar{n}+1)\\
				L_2=\proj{4}{2};&\quad \gamma_2=\gamma(\bar{n})\\
				L_3=\proj{1}{3};&\quad \gamma_3=\gamma(\bar{n}+1)\\
				L_4=\proj{3}{1};&\quad \gamma_4=\gamma(\bar{n}).
			\end{align}
		\end{subequations}
		Considering a discrete time evolution model, we may evaluate the evolution of the state for a time interval of $\Delta t$, by considering a Kraus operator sum representation of the form
		\begin{align}
			\rho(t+\Delta t) = \sum_{k} M_k(\Delta t) \rho(t) M_k^\dagger(\Delta t)
		\end{align}
		We define the Kraus operators (based on the Lindblad master equation) as
		\begin{align}
			M_0 = \mathbb{I}_{\mathrm{sys}}+(K-iH_{\mathrm{sys}})\Delta t; \qquad M_k=\sqrt{\gamma_k \Delta t} L_k; k\neq0
		\end{align}
		Where the operator $K$ is defined as $K=-\frac{1}{2}\sum_i \gamma_i L_i^\dagger L_i$. For the evolution prescribed by Eq.~\eqref{eq:master_Lindblad}, we have 
		\begin{align}
			\begin{split}
				K&=-\frac{1}{2}\left( \gamma(\bar{n}+1) \outprod{4} + \gamma(\bar{n}+1) \outprod{3} + \gamma\bar{n} \outprod{2} + \gamma\bar{n} \outprod{1}\right).
			\end{split}
		\end{align}
		We then have the following expression for $M_0$,
		\begin{align}
			M_0= \mathbb{I}_{\mathrm{sys}} -\frac{1}{2} \left[ \left(\gamma(\bar{n}+1)+\hbar\omega'_A\right)(\outprod{4}+\outprod{3}) + \left(\gamma(\bar{n})-\hbar\omega'_A\right)(\outprod{2}+\outprod{1}) \right]\Delta t
		\end{align}
		and the other Kraus operators defined as, 
		\begin{subequations}
			\begin{align}
				M_1&=\sqrt{\gamma(\bar{n}+1)\Delta t}\proj{2}{4}\\
				M_2&=\sqrt{\gamma(\bar{n})\Delta t}\proj{4}{2}\\
				M_3&=\sqrt{\gamma(\bar{n}+1)\Delta t}\proj{1}{3}\\
				M_4&=\sqrt{\gamma(\bar{n})\Delta t}\proj{3}{1}
			\end{align}
		\end{subequations}
		It is quite straightforward to verify that the Kraus operators satisfy $\sum_{k=0}^4 M_k^\dagger M_k=\mathbb{I}$. We first note that the operator $ K$ is self-adjoint,
		\begin{align}
			\begin{split}
				K^\dagger&=-\frac{1}{2}\left[\sum_i \gamma_i L_i^\dagger L_i\right]^\dagger=-\frac{1}{2}[\sum_i \gamma_i L_i^\dagger (L_i^\dagger)^\dagger=-\frac{1}{2}\sum_i \gamma_i L_i^\dagger L_i=K
			\end{split}
		\end{align}
		Evaluating $M^\dagger_0 M_0$ we get
		\begin{subequations}
			\begin{align}
				M^\dagger_0 M_0 & = \left( \mathbb{I}_{\mathrm{sys}}+(K-iH_{\mathrm{sys}})\Delta t \right)^\dagger \left( \mathbb{I}_{\mathrm{sys}}+(K-iH_{\mathrm{sys}})\Delta t\right)\\
				&=\mathbb{I} + (K^\dagger +K)\Delta t  +i(H_{\mathrm{sys}}^\dagger-H_{\mathrm{sys}})\Delta t + (K^\dagger+i H_{\mathrm{sys}}^\dagger) (K-i H_{\mathrm{sys}})(\Delta t)^2\\
				&= \mathbb{I} + 2 K \Delta t \qquad \quad\text{(ignoring terms of order $(\Delta t)^2$)}
			\end{align}
		\end{subequations}
		We note that $\sum_{k=1}^4 M^\dagger_k M_k= \sum_{i=1,4}\gamma_i L_i^\dagger L_i= -2K \Delta t$. Hence $\sum_{k=0}^4 M_k^\dagger M_k=\mathbb{I}$ is clearly satisfied.
		
	}
	
	{ 
		\section{Analysis of Heavier Group IV Vacancy Centers}
		\label{app:heavier_vacancy}
		
		\begin{figure}[ht!]
			\centering
			\includegraphics[width=0.6\linewidth]{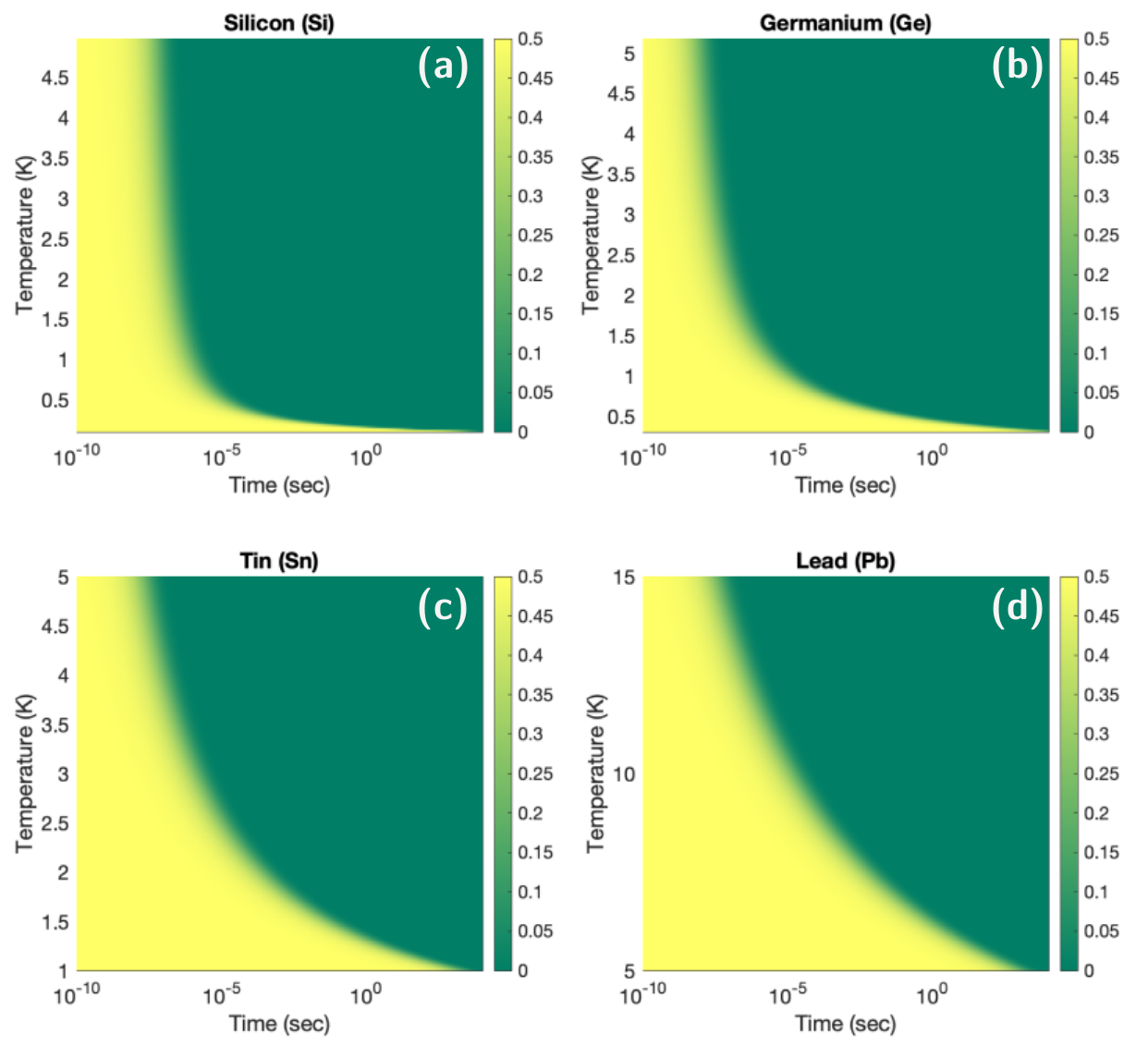}
			\caption{ Evolution of the single spin state coherence   $ \braket{\downarrow\!|\rho(t)|\!\uparrow}= \braket{\uparrow\!|\rho(t)|\!\downarrow} $  initialized in the  $ \ket{\psi(t=0)} =(\ket{1}+\ket{2})/\sqrt{2}$ state. We look at dependence of $ \braket{\downarrow\!|\rho(t)|\!\uparrow} $ at various values of bath temperature $ (T) $ and maximum time for (a) silicon (Si), (b) germanium (Ge), (c) tin (Sn), and (d) lead (Pb) defects. }
			\label{fig:sing_qbit_coh_dplot}
		\end{figure}
		
		Even though many of the latest studies with color centers in diamond have been performed with the silicon vacancy (SiV)~\cite{Sipahigil2016-de,Bhaskar2020-kv,Stas2022-yw}, their scalability is limited due to their stringent operating conditions (e.g., $\approx 100$ mK operating temperature for use as a useful quantum memory). Heavier group IV vacancy centers such as the germanium (Ge)~\cite{Bhaskar2017-da,Iwasaki2015-kh,Wan2020-op}, tin (Sn)~\cite{Debroux2021-mw,Trusheim2020-ko}, and lead (Pb)~\cite{Trusheim2019-ck,Wang2021-qp} vacancies in diamond are being explored in several parallel efforts across the world for their reduced susceptibility to phonon-induced decoherence as a consequence of increased ground state manifold orbital splitting. The following splitting values have been calculated and experimentally measured:
		
		\begin{table}[h]
			\centering
			\begin{tabular}{p{2cm} p{3cm} p{3cm}}
				\toprule
				Defect Atom&  Ground State Orbital Splitting (GHz)& Typical Operating Temperatures (K)\\
				\midrule Si&  50& 0.1\\
				Ge&  181& 0.4\\
				Sn&  640& 1\\
				Pb&  3750& $>$4\\
				\bottomrule
			\end{tabular}
			\caption{Ground state splitting and operating temperatures for heavier vacancy centers in diamond.}
			\label{tab:vac_params}
		\end{table}
		
		We extend the analysis from Sec.~\ref{sec:sing_spin_decoh} for the heavier vacancy centers in Fig.~\ref{fig:sing_qbit_coh_dplot} to obtain the single spin state coherence   $ \braket{\downarrow\!|\rho(t)|\!\uparrow}= \braket{\uparrow\!|\rho(t)|\!\downarrow} $  as function of temperature and time for an initial qubit state $ \ket{\psi(t=0)} =(\ket{1}+\ket{2})/\sqrt{2}$. We plot the extracted coherence decay time constant in Fig.~\ref{fig:sing_qbit_coh_lines} - we note that equivalent coherence times are observed for higher temperatures, as the defect center's spin-orbit splitting becomes larger. The same holds true for two-qubit entanglement decay (similar to Fig.~\ref{fig:ent_spin_decohere}) or over a network (from Fig.~\ref{fig:singleRail_decohere} and~\ref{fig:dualRail_decohere}). Overall, utilizing a heavier vacancy center species would potentially allow for less stringent operating conditions. Additional operational difficulties (such as in sample fabrication, faithful qubit manipulation and high-efficiency photon collection) may arise in using these other emitters; however they are out of the scope of this present article.

		\begin{figure}[ht!]
			\centering
			\includegraphics[width=0.8\linewidth]{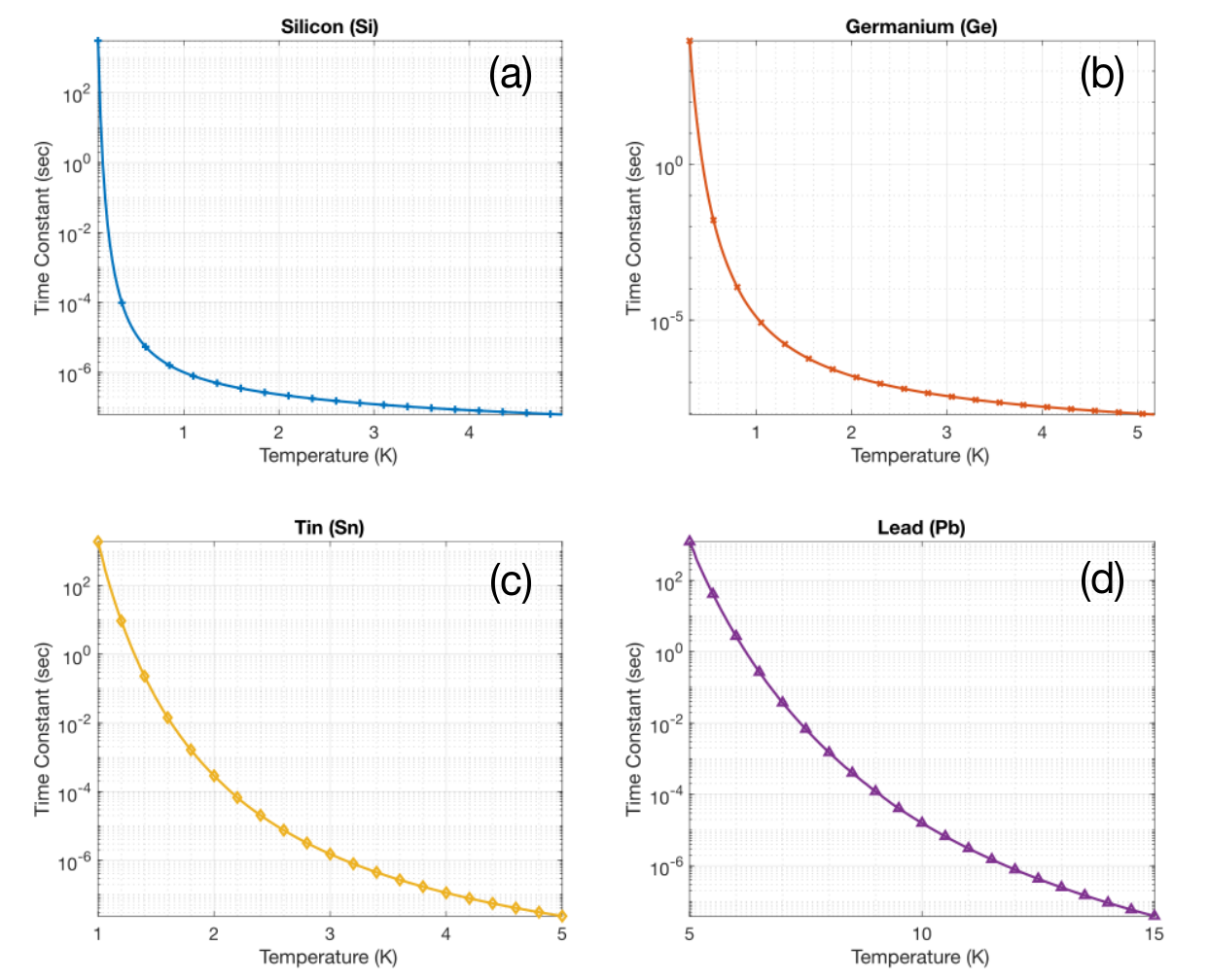}
			\caption{ Extracted decay time constant by fitting $ \braket{\downarrow\!|\rho(t)|\!\uparrow}= \braket{\uparrow\!|\rho(t)|\!\downarrow} $  to the expected relation  $ \braket{\uparrow\!|\rho(t)|\!\downarrow}  = 0.5\times \exp (-t/\tau_{C,1}) $ for (a) silicon (Si), (b) germanium (Ge), (c) tin (Sn), and (d) lead (Pb) defects. }
			\label{fig:sing_qbit_coh_lines}
		\end{figure}
	}

	\twocolumngrid
	
	\bibliography{biblio_decoherence}

\end{document}